\documentclass[10pt,a4paper,headings=normal,oneside,numbers=noendperiod]{scrartcl}

\usepackage[ngerman,english]{babel}
\usepackage{lmodern}
\usepackage[authoryear]{natbib}
\usepackage{array}
\usepackage{longtable}
\usepackage{booktabs}
\usepackage{graphicx}
\usepackage{subfigure}
\usepackage{amsmath}
\usepackage{amssymb}
\usepackage{sistyle}
\usepackage{nicefrac}
\usepackage{paralist}
\usepackage{setspace}

\newcommand{\myfootnote}[1]{%
\renewcommand{\thefootnote}{}%
\footnotetext{#1}%
\renewcommand{\thefootnote}{\arabic{footnote}}%
}

\setkomafont{disposition}{\bfseries\rmfamily}

\begin{document}

\begin{onehalfspace}
\noindent
{\textbf{\Large Investigations of the coordinates in Ptolemy's\\ \textit{Geographike Hyphegesis} Book 8}}
\\
\end{onehalfspace}
\noindent
Christian Marx$^*$\myfootnote{$^*$C. Marx, Gropiusstra{\ss}e 6, 13357 Berlin, Germany; e-mail: ch.marx@gmx.net.}
\myfootnote{\;\,Author generated postprint; published in Archive for History of Exact Sciences (2012) 66: 531--555;}
\myfootnote{\;\,DOI 10.1007/s00407-012-0102-0.}
\\\\\\\\
\noindent
\textbf{Abstract}\quad In Book 8 of his \textit{Geographike Hyphegesis} Ptolemy gives coordinates for ca. 360 so-called noteworthy cities. These coordinates are the time difference to \textit{Alexandria}, the length of the longest day, and partly the ecliptic distance from the summer solstice.
The supposable original conversions between the coordinates in Book 8 and the geographical coordinates in the location catalogue of Books 2--7 including the underlying parameters and tabulations are here reconstructed. The results document the differences between the $\mathrm{\Omega}$- and $\mathrm{\Xi}$-recension.
The known difference in the longitude of \textit{Alexandria} underlying the conversion of the longitudes is examined more closely.
For the ecliptic distances from the summer solstice of the $\mathrm{\Omega}$-recension it is revealed that they were originally computed by means of a so far undiscovered approximate, linear conversion. Further it is shown that the lengths of the longest day could be based on a linear interpolation of the data in the \textit{Mathematike Syntaxis} 2.6.
\\\\
\noindent
\textbf{Keywords}\quad Ptolemaios, \textit{Geographike Hyphegesis} Book 8, Conversion of coordinates, Longitude, Latitude, Length of the longest day, Ecliptic distance

\section{Introduction} \label{sec:intro}

\begin{figure}
\begin{center}
\includegraphics{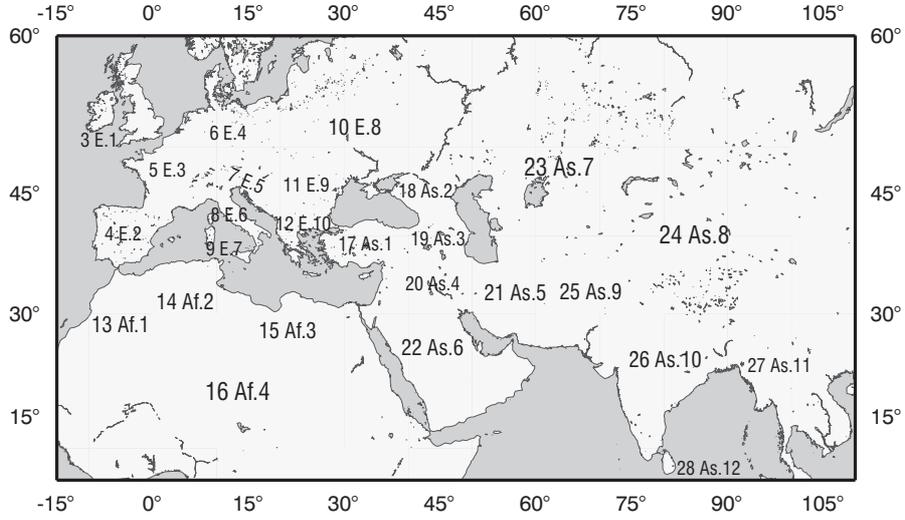}
\caption{Overview of the chapters in Book 8 (E.: Europe, Af.: Africa, As.: Asia)}
\label{fig:kapitel}
\end{center}
\end{figure}

Ptolemy's \textit{Geographike Hyphegesis} (GH/Geography) compiled in about 150 AD contains a location catalogue in Books 2--7 with about 6,300 ancient localities and their positions expressed by geographical longitude $\mathrm{\Lambda}$ and latitude $\mathrm{\Phi}$. 360 of these localities are additionally listed with coordinates in Book 8 of the Geography. In contrast to the location catalogue, the positions in Book 8 are expressed by means of
\begin{compactitem} [--]
\item the time difference $A$ (in hours) from the location to \textit{Alexandria},
\item the length of the longest day $M$ (in hours) at the location,
\item the ecliptic distance $S$ (in arc degree) of the sun from the summer solstice at the time when the sun reaches the zenith (for locations between the tropics).
\end{compactitem}
The localities in Book 8 are selected cities, the so-called \textit{poleis episemoi} (``noteworthy cities'').
They are arranged in 26 chapters; each chapter is related to one of the maps Ptolemy divided the \textit{Oikoumene} into (the inhabited world known to the Greeks and Romans).
Figure \ref{fig:kapitel} shows a general view of the sites of the single maps \citep[see also][]{stu06}.

Since the positions of the \textit{poleis episemoi} are given in the location catalogue and in Book 8, a comparison of the positions of both sources is possible after a conversion of coordinates.
The \textit{poleis episemoi} also appear in another of Ptolemy's works, the \textit{Procheiroi Kanones} (Handy Tables; hereinafter denoted Canon). Unlike Book 8 the coordinates are given there in form of geographical coordinates as in the location catalogue of the Geography.

A conversion of the $A$- and $M$-data in Book 8 into geographical coordinates can be found in  \citet{cun23} and in \citet{stu09}. The latter contains an edition of the Canon with a compilation of the corresponding geographical coordinates of the GH~2--7 and the converted values $\mathrm{\Lambda}(A)$ and $\mathrm{\Phi}(M)$ on the basis of Book 8. \citeauthor{cun23} examines four maps of Europe (GH~8.5--8). He supposes that $A$ and $M$ in Book 8 originate from the geographical coordinates in Books 2--7. By his assumptions about the  precision of $A$ and $M$ as well as about tabulations  underlying $A$ and $M$  he  makes an attempt  to explain the coordinate values in Book 8 (see Sects.~\ref{sec:alex} and \ref{sec:laengster_tag}).
\citet[p. 64\,ff.]{hon29} assumes that the data of the \textit{poleis episemoi}  in Book 8 as well as in Books 2--7 were derived from works by a predecessor of Ptolemy.

In the following, the conversions of coordinates between the location catalogue and Book 8 are investigated.
The underlying  calculation methods and their parameters are worked out as well as the  underlying tabulations.
Deviations between the coordinates in Book 8 and in the location catalogue are explained as comprehensively as possible by the resolution (precision) of the coordinates.
As source for the data of the Geography, the edition of \citet{stu06} was used (for changes on the data see Appendix). It contains the coordinates of the two main recensions of the Geography, the $\mathrm{\Omega}$- and $\mathrm{\Xi}$-recension. The latter one has been handed down only by the X-manuscript; its location catalogue ends in Book 5, Chapter 13, so that a direct comparison with the data in Book 8 is not possible. Both $\mathrm{\Omega}$  and X are considered in the present work.
\citet[p. 44]{stu06} state that many $M$-values of X seem to be based on the coordinates of the Canon.
The relation between the coordinates in Book 8 and the Canon is only broached here. The coordinates used are based on the edition of the Canon by \citet{stu09}.

\section{Time difference to \textit{Alexandria}} \label{sec:alex}

The conversion between the time difference to Alexandria $A$ (in hours) and the longitude $\mathrm{\Lambda}$ (in degree) is
\begin{equation} \label{eqn:A2L}
\mathrm{\Lambda} = A \ \; 15 \nicefrac{\arcdeg}{\mathrm{h}} + \mathrm{\Lambda_A}
\end{equation}
with the longitude of Alexandria $\mathrm{\Lambda_A}$ (in degree).
Ptolemy gives two different values for $\mathrm{\Lambda_A}$ in the Geography, in the location catalogue $\ang{60;30}=:\mathrm{\bar{\Lambda}_{A}}$ (4.5.9) and in Book 8 $ 4\,\mathrm{h}=:\mathrm{\tilde{\Lambda}_{A}}$ (8.15.10; the distance between \textit{Alexandria} and the zero meridian); in GH~7.5.14, he gives both values. The difference between both parameters is $\mathrm{\Delta \Lambda_A}:=\mathrm{\bar{\Lambda}_{A}}-\mathrm{\tilde{\Lambda}_{A}}=\ang{60;30}-60\arcdeg=30\arcmin=2$\,min.

\citet[p. 101\,ff.]{cun23} investigates the coordinates of the 3rd to 6th map in GH~8.5--8 (Europe, 42 localities). He states that both $\ang{60;30}$ and $\ang{60;00}$ underlie the time differences $A_i$ ($i=1,2,\ldots,$number of coordinates $n$) and that it is not possible to determine the $\mathrm{\Lambda_A}$ the coordinates are based on in all cases.
His explanation for different $\mathrm{\Lambda_A}$ is that Ptolemy upgraded the value $\ang{60;30}$ to $\ang{60;00}$ and added the updated $A_i$ in a further column besides the values underlying $\ang{60;30}$; later copyists chose one of the values.
\citet[p. 44]{stu06} mention that $60\arcdeg$ primarily underlies the data of $\mathrm{\Omega}$ as well as $\ang{60;30}$ the data of X.

\subsection{Conversion of coordinates and accuracy}

Ptolemy explicitly declares in GH~8.2.1 that the time differences $A_i$ in Book 8 are approximate values; thus, equality after a conversion between the data of Books 2--7 and Book 8 is not to be expected.
For the following conversions of coordinates, the difference $\mathrm{\Delta} y$  between the conversion $y(x)$ of a coordinate $x$ and the given coordinate value $y$ is defined as
\begin{equation} \label{eqn:diff}
\mathrm{\Delta} y = y - y(x) \;.
\end{equation}
Then $\mathrm{\Delta} y$ is the error of $y$ with respect to $y(x)$.

\begin{figure}
\centering
\subfigure{
\includegraphics[width=5.6cm]{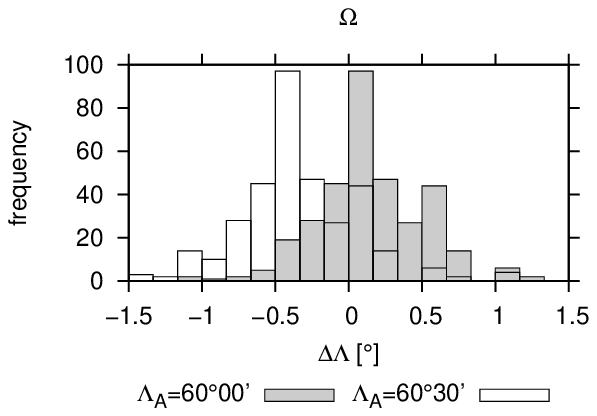}
}
\subfigure{
\includegraphics[width=5.6cm]{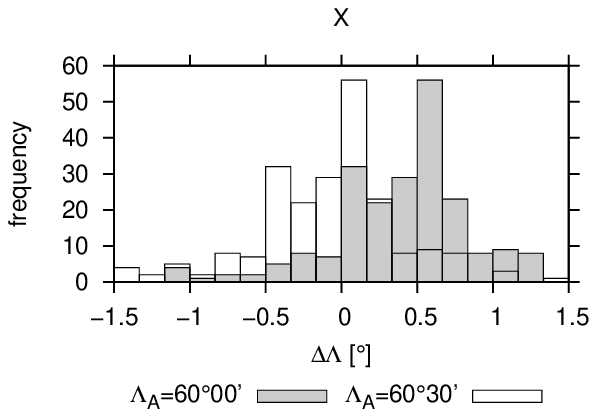}
}
\caption{Frequency of differences $\mathrm{\Delta \Lambda_i}=\mathrm{\Lambda}_i-\mathrm{\Lambda}(A_i)$ between the longitudes $\mathrm{\Lambda}_i$ of the location catalogue and the longitudes $\mathrm{\Lambda}(A_i)$ computed from the time differences $A_i$ to \textit{Alexandria} of Book 8 by means of Eq.~(\ref{eqn:A2L}) for the $\mathrm{\Omega}$-recension (\textit{left}) and the X-manuscript (\textit{right}); $\mathrm{\Lambda_A}$ is the longitude of \textit{Alexandria} underlying the conversion}
\label{fig:diff_l}
\end{figure}

Both $\mathrm{\bar{\Lambda}_{A}}$ and $\mathrm{\tilde{\Lambda}_{A}}$ were used here for the conversion of coordinates.
For the conversion $A\mapsto \mathrm{\Lambda}$ by Eq.~(\ref{eqn:A2L}) Fig.~\ref{fig:diff_l} shows the frequencies of differences $\mathrm{\Delta \Lambda}_i=\mathrm{\Lambda}_i-\mathrm{\Lambda}(A_i)$ lying  within specified intervals (interval width $10\arcmin$).
For $\mathrm{\Omega}$ the conversion with $\mathrm{\tilde{\Lambda}_{A}}$ gives a maximal frequency near zero and with $\mathrm{\bar{\Lambda}_{A}}$ near $-30\arcmin$. That is the negative systematic difference $\mathrm{\Delta \Lambda_A}$ and  indicates that the majority of the coordinates were originally converted with $\mathrm{\tilde{\Lambda}_{A}}=60\arcdeg$.
In X the conditions are converse so that $\mathrm{\bar{\Lambda}_{A}}=\ang{60;30}$ must be the main parameter of the original conversion.
The frequency distribution is bimodal; the distance of the lower second peak from the main peak is $30\arcmin$. Hence, the second peak is probably caused by  $A_i$ which are based on the corresponding other $\mathrm{\Lambda_A}$.

The size of the differences $\mathrm{\Delta} \mathrm{\Lambda}_i$ can mainly be explained by a original conversion $\mathrm{\Lambda}\mapsto A$ in conjunction with a rounding of $A$.
Apart from a few exceptions, the largest denominator of the fractions of the $A_i$ is $30$ ($\nicefrac{1}{30}\,\mathrm{h}=2\,\mathrm{min}$).
\citet[p. 108]{cun23} assumes an original precision of $\nicefrac{1}{12}\,\mathrm{h}=5\,\mathrm{min}$ ($=\ang{1;15}$).\footnote{\citet{hon29}, who discusses \citeauthor{cun23}' work, does not dissent.}
In general, if a number is rounded with a precision (resolution) $a=1/d$, the expected maximal rounding error is
\begin{equation} \label{eqn:rundung}
e_\mathrm{max} = \pm \frac{a}{2} = \pm \frac{1}{2d} \; .
\end{equation}
For $a=5\,\mathrm{min}$ $e_\mathrm{max}$ is $2.5\,\mathrm{min}\approx \nicefrac{2}{3}\arcdeg$. This agrees with that part of the tail of the frequency distribution which is not extended by the systematic difference $\mathrm{\Delta \Lambda_A}$ (see Fig.~\ref{fig:diff_l}).

\citet[p. 108]{cun23} gives a tabulation of $A$- and $\mathrm{\Lambda}$-values with a step size of $\mathrm{\Delta}A=5$\,min ($\mathrm{\Delta}\mathrm{\Lambda}=1\nicefrac{1}{4}\arcdeg$) by which $A$-values may have been determined. \citeauthor{cun23} supposes that in the case of a $\mathrm{\Lambda}_i$ which does not equal a tabulated $\mathrm{\Lambda}$ the  nearest tabulated $\mathrm{\Lambda}$ was chosen. In the end \citeauthor{cun23} cannot explain several $A_i$ by means of the tabulation.
The use of a tabulation described by \citeauthor{cun23} was applied here to the entire $A$-data of Book 8. A tabulation was generated for $\mathrm{\bar{\Lambda}_{A}}$ and $\mathrm{\tilde{\Lambda}_{A}}$. In the case of $\mathrm{\Omega}$ 61\% and in the case of X 51\% of the $A$-values determined by one of both tabulations equal the $A_i$ given in Book 8. Furthermore a tabulation with $\mathrm{\Delta}A=2.5$\,min was tested; the result are 61\% and 55\%. That does not argue for the use of a tabulation. \citeauthor{cun23} gives the example of \textit{Lugdunum} (8.5.5) with $\mathrm{\Lambda}=\ang{23;15}$ and $A=-2^\mathrm{h}30^\mathrm{m}$. The nearest tabulated $\mathrm{\Lambda}$ is $23\arcdeg$ so that $A$ becomes $-2^\mathrm{h}30^\mathrm{m}$. However, this might also be the result of Eq.~(\ref{eqn:A2L}) and a rounding: $A(\mathrm{\Lambda})=-2^\mathrm{h}29^\mathrm{m}$.

\subsection{Determination of $\mathrm{\Lambda_A}$ and the resolution}

Considering the single chapters of Book 8, there are chapters whose longitudes are converted with one of both  $\mathrm{\Lambda}_\mathrm{A}$ mainly; in other cases, the underlying $\mathrm{\Lambda}_\mathrm{A}$ cannot be recognized reliably.
The question of the $\mathrm{\Lambda}_\mathrm{A}$ underlying a single $A$ led to an attempt  to determine  $\mathrm{\Lambda}_\mathrm{A}$ for each locality by means of the following method. $\mathrm{\Lambda}_i$ is converted with  $\mathrm{\bar{\Lambda}}_\mathrm{A}$ to $\bar{A}(\mathrm{\Lambda}_i)$ and with $\mathrm{\tilde{\Lambda}}_\mathrm{A}$ to $\tilde{A}(\mathrm{\Lambda}_i)$ by Eq.~(\ref{eqn:A2L}). $\bar{A}_i$ and $\tilde{A}_i$ are rounded with different precisions $a$ to determine the apparent resolution of the given $A_i$.
In general terms, the two coordinates $x$ and $y$ of different types are given as well as the converted coordinate $y(x)$. Furthermore a main resolution $a_\mathrm{s}$ is set. The procedure applied is then\\
Procedure \texttt{ROUND}:
\begin{compactenum} [(1)]
\item Round $y(x)$ to the precision $a_\mathrm{s}$; if $\mathrm{\Delta}y = y-y(x)= 0$ end.
\item For each of the precisions $a<a_\mathrm{s}$ downwards: round $y(\mathrm{x})$; if $\mathrm{\Delta}y = 0$ end.
\item For each of the precisions $a>a_\mathrm{s}$ upwards: round $y(\mathrm{x})$; if $\mathrm{\Delta}y = 0$ end.
\end{compactenum}
If $\mathrm{\Delta}y$ becomes $0$, the apparent resolution of $y$ is found; if $\mathrm{\Delta}y$ does not become $0$ until the end,  an erroneous $y$ or $x$ can be assumed.

Procedure \texttt{ROUND} was applied to the $\bar{A}_i$ and $\tilde{A}_i$. The results of a comparison with the given $A_i$ were valued as follows:

\begin{compactenum} [(1)]
\item rounded $\bar{A}_i=A_i$ or/and rounded $\tilde{A}_i=A_i$:
    \begin{compactenum} [(a)]
    \item only rounded $\bar{A}_i=A_i$: $A_i$ is apparently based on $\mathrm{\bar{\Lambda}_{A}}$;
    \item only rounded $\tilde{A}_i=A_i$: $A_i$ is apparently based on $\mathrm{\tilde{\Lambda}_{A}}$;
    \item rounded $\bar{A}_i=A_i$ and rounded $\tilde{A}_i=A_i$: $\mathrm{\Delta}\bar{A}_i=A_i-\bar{A}_i$, $\mathrm{\Delta}\tilde{A}_i=A_i-\tilde{A}_i$
        \begin{compactenum} [(i)]
        \item $|\mathrm{\Delta} \bar{A}_i| < |\mathrm{\Delta} \tilde{A}_i|$: $A_i$ is possibly  based on $\mathrm{\bar{\Lambda}_{A}}$;
        \item $|\mathrm{\Delta} \bar{A}_i| > |\mathrm{\Delta} \tilde{A}_i|$: $A_i$ is possibly  based on $\mathrm{\tilde{\Lambda}_{A}}$;
        \item $|\mathrm{\Delta} \bar{A}_i| = |\mathrm{\Delta} \tilde{A}_i|$: no decision possible;
        \end{compactenum}
    \end{compactenum}
\item rounded $\bar{A}_i\neq A_i$ and rounded $\tilde{A}_i\neq A_i$: no decision possible, maybe a corrupted $A_i$-value.
\end{compactenum}

The possible resolutions for Procedure \texttt{ROUND} were chosen according to the denominators of the fractions which occur in Book 8. The resolutions used for $\mathrm{\Omega}$ are (in h): \nicefrac{1}{360}, \nicefrac{1}{60}, \nicefrac{1}{30}, \nicefrac{1}{24}, \nicefrac{1}{15}, \nicefrac{1}{12}, \nicefrac{1}{10}, \nicefrac{1}{8}, \nicefrac{1}{7}, \nicefrac{1}{6}, and  \nicefrac{1}{5}\footnote{The denominator 360 does not really occur but was introduced because of the occurrence of the sum $\nicefrac{1}{8}+\nicefrac{1}{90}=\nicefrac{49}{360}$ \citep[see][p. 883, n. 65]{stu06}.}. For X \nicefrac{1}{60} and \nicefrac{1}{7} were not used, and \nicefrac{1}{20} and \nicefrac{1}{9} were added.
The main resolution  was set to $a_\mathrm{s}=\nicefrac{1}{12}$\,h \citep[following][]{cun23} and for a further run to $a_\mathrm{s}=\nicefrac{1}{24}$\,h.
The results for the portions of apparent resolutions are:
\begin{compactitem} [--]
\item $a_\mathrm{s}=\nicefrac{1}{12}$\,h:
    \begin{compactitem} [--]
    \item $\mathrm{\Omega}$: $a=a_\mathrm{s}$: 61\%, $a<a_\mathrm{s}$: 34\%, $a>a_\mathrm{s}$: 4\%, no fitting rounding: 1\%;
    \item X: $a=a_\mathrm{s}$: 51\%,  $a<a_\mathrm{s}$: 36\%, $a>a_\mathrm{s}$: 9\%, no fitting rounding: 4\%;
    \end{compactitem}
\item $a_\mathrm{s}=\nicefrac{1}{24}$\,h:
    \begin{compactitem} [--]
    \item $\mathrm{\Omega}$: $a=a_\mathrm{s}$: 61\%, $a<a_\mathrm{s}$: 24\%, $a>a_\mathrm{s}$: 14\%, no fitting rounding: 1\%;
    \item X: $a=a_\mathrm{s}$: 55\%, $a<a_\mathrm{s}$: 23\%, $a>a_\mathrm{s}$: 19\%, no fitting rounding: 4\%.
    \end{compactitem}
\end{compactitem}

The result indicates that $a_\mathrm{s}=\nicefrac{1}{24}$\,h is the main resolution, but also $a_\mathrm{s}=\nicefrac{1}{12}$\,h comes into consideration.
The result of Procedure \texttt{ROUND} applied with  $a_\mathrm{s}=\nicefrac{1}{24}$\,h shows that ca. 80\% of the localities have only small absolute differences $<0.5 \cdot \nicefrac{1}{24}\,\mathrm{h}=\nicefrac{1}{48}\,\mathrm{h}=1.25\,\mathrm{min}$ (cf. Eq.~(\ref{eqn:rundung})) between $A_i$ and $A(\mathrm{\Lambda}_i)$ converted by Eq.~(\ref{eqn:A2L}).
Further, almost all $A_i$ can be explained by conversion based on either $\mathrm{\bar{\Lambda}_A}$ or $\mathrm{\tilde{\Lambda}_A}$ and a rounding. Hence, the vast majority of the differences between the given $A_i$ and the converted $A(\mathrm{\Lambda}_i)$ are explicable by the precision of the $A_i$. Consequently, the location catalogue and Book 8 show a high level of interdependence.

The determined frequencies of $\mathrm{\bar{\Lambda}_A}$ and $\mathrm{\tilde{\Lambda}_A}$ are:
\begin{compactitem} [--]
\item $\mathrm{\Omega}$: $\mathrm{\bar{\Lambda}_A}$: 107 (31\%), $\mathrm{\tilde{\Lambda}_A}$: 209 (61\%), $\mathrm{\bar{\Lambda}_A}$ or $\mathrm{\tilde{\Lambda}_A}$: 25 (7\%), no fitting rounding: 3 (1\%);
\item X: $\mathrm{\bar{\Lambda}_A}$ 137 (62\%), $\mathrm{\tilde{\Lambda}_A}$: 63 (29\%), $\mathrm{\bar{\Lambda}_A}$ or $\mathrm{\tilde{\Lambda}_A}$: 11 (5\%), no fitting rounding: 9 (4\%).
\end{compactitem}
This again reveals that $\mathrm{\bar{\Lambda}_A}$ dominates in X and $\mathrm{\tilde{\Lambda}_A}$ in $\mathrm{\Omega}$ (with a ratio of ca. 2:1).

\begin{figure}
\begin{center}
\includegraphics[width=120mm]{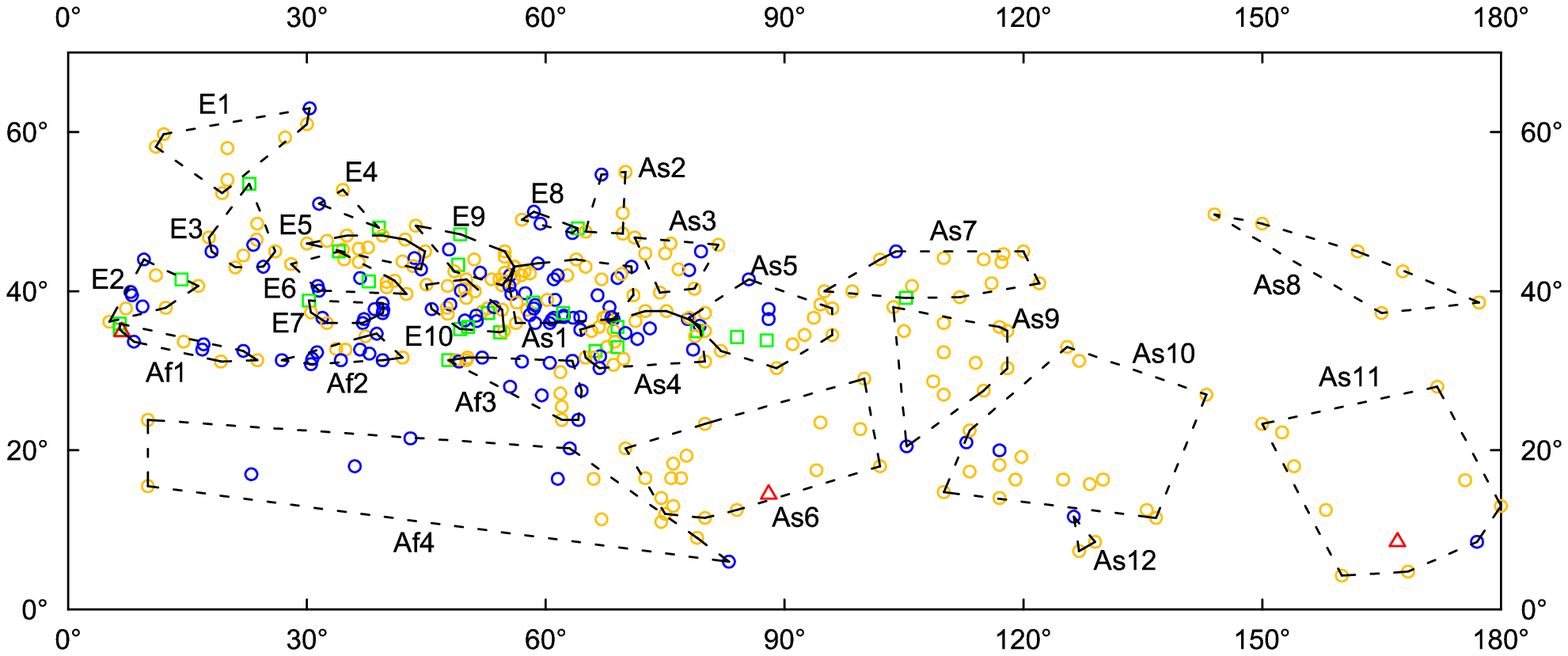}
\caption{Classification of the localities in Book 8 of the $\mathrm{\Omega}$-recension according to the longitude $\mathrm{\Lambda_{A}}$ of \textit{Alexandria} which underlies a conversion of the longitudes $\mathrm{\Lambda}_i$ of the location catalogue into the time differences $A_i$ to \textit{Alexandria} of Book 8; \textit{blue}/\textit{circle}: $\mathrm{\bar{\Lambda}_{A}}=\ang{60;30}$, \textit{yellow}/\textit{circle}: $\mathrm{\tilde{\Lambda}_{A}}=\ang{60;00}$, \textit{green}/\textit{square}: $\mathrm{\bar{\Lambda}_{A}}$ or $\mathrm{\tilde{\Lambda}_{A}}$, \textit{red}/\textit{triangle}: $A$ does not fit with $\mathrm{\Lambda}$}
\label{fig:orte_la_o}
\end{center}
\end{figure}

\begin{figure}
\begin{center}
\includegraphics[width=65mm]{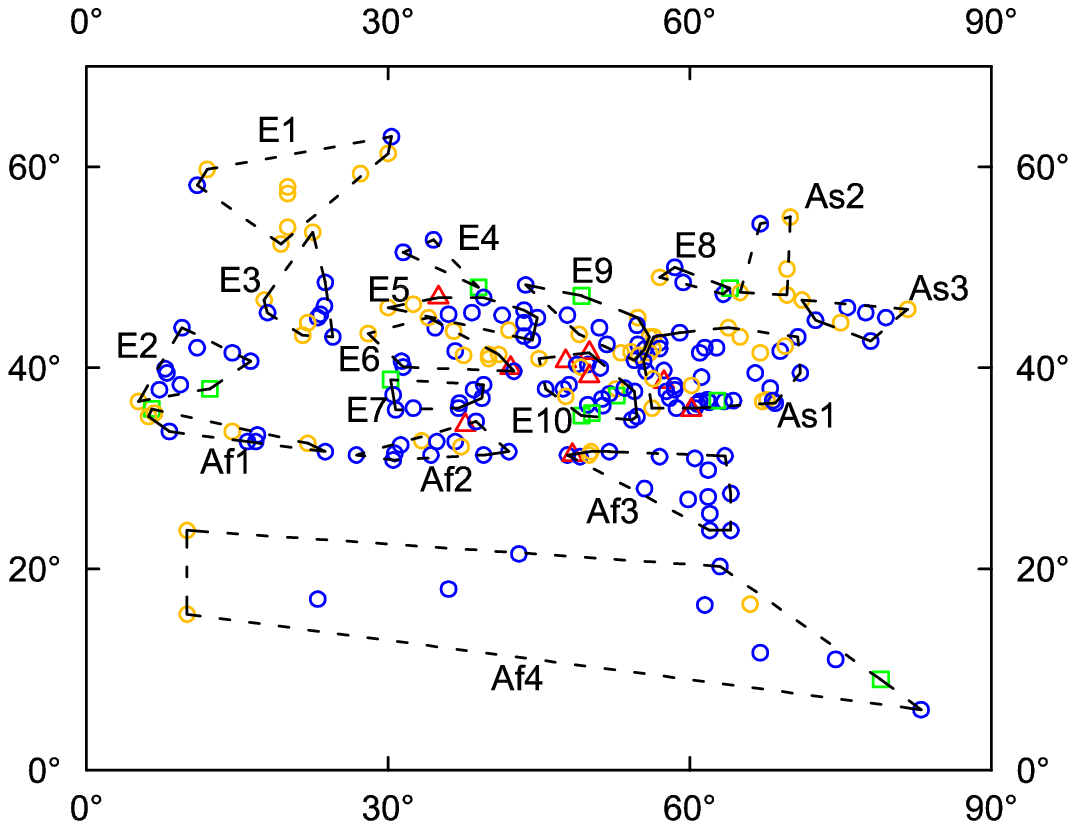}
\caption{Classification of the localities in Book 8 of the X-manuscript according to the longitude $\mathrm{\Lambda_{A}}$ of \textit{Alexandria} which underlies a conversion of the longitudes $\mathrm{\Lambda}_i$ of the location catalogue into the time differences $A_i$ to \textit{Alexandria} of Book 8;  \textit{blue}/\textit{circle}: $\mathrm{\bar{\Lambda}_{A}}=\ang{60;30}$, \textit{yellow}/\textit{circle}: $\mathrm{\tilde{\Lambda}_{A}}=\ang{60;00}$, \textit{green}/\textit{square}: $\mathrm{\bar{\Lambda}_{A}}$ or $\mathrm{\tilde{\Lambda}_{A}}$, \textit{red}/\textit{triangle}: $A$ does not fit with $\mathrm{\Lambda}$}
\label{fig:orte_la_x}
\end{center}
\end{figure}

Figures \ref{fig:orte_la_o} and \ref{fig:orte_la_x} show the result of the classification for each locality in the Ptolemaic coordinate system ($\mathrm{\bar{\Lambda}_A}$: blue/circle, $\mathrm{\tilde{\Lambda}_A}$: yellow/circle, $\mathrm{\bar{\Lambda}_A}$ or $\mathrm{\tilde{\Lambda}_A}$: green/square, no fitting rounding: red/triangle).
In $\mathrm{\Omega}$ the dominant $\mathrm{\tilde{\Lambda}_{A}}$ occurs nearly exclusively in the east, and in the west also $\mathrm{\bar{\Lambda}_{A}}$ occurs often.
There are in $\mathrm{\Omega}$ 13 chapters in which $\mathrm{\tilde{\Lambda}_{A}}$ dominates: Europe 1, 3, 5, 6, 9, Asia 3, 5--11; in 2 chapters $\mathrm{\bar{\Lambda}_{A}}$ dominates: Europe 7, Africa 2.
In X $\mathrm{\bar{\Lambda}_{A}}$ dominates in 8 chapters: Europe 2, 5, 7, 10, Africa 2--4, Asia 1;   $\mathrm{\tilde{\Lambda}_{A}}$ dominates in 2 chapters: Europe 1, Asia 2.

\subsection{Comparison of $\mathrm{\Omega}$ and X}

Finally, the $A_i$ of $\mathrm{\Omega}$ and X were compared. 343 values are equal and 143 values have differences, thereof 16 localities also have differences in $\mathrm{\Lambda}$ in the location catalogue so that $143-16=127$  $A_i$ remain with actual differences between $\mathrm{\Omega}$ and X (37\%).
Besides the expected systematic difference of $\mathrm{\Delta \Lambda_A} = 30\arcmin = 2\,\mathrm{min}$ also other differences occur comparably often: $2.5\,\mathrm{min}=\nicefrac{1}{24}$\,h, $3\,\mathrm{min}=\nicefrac{1}{20}$\,h, $4\,\mathrm{min}=\nicefrac{1}{15}$\,h, and $5\,\mathrm{min}=\nicefrac{1}{12}$\,h. This is explicable by the precision of the $A_i$. As an example \textit{Clunia} (8.4.5) is considered: In $\mathrm{\Omega}$ $A$ is $-3^\mathrm{h}15^\mathrm{m}$, that is $\mathrm{\Lambda}=\ang{11;00}$ converted with $\mathrm{\tilde{\Lambda}_{A}}=60\arcdeg$ to $A=-3^\mathrm{h}16^\mathrm{m}$ and rounded to the nearest \nicefrac{1}{12}\,h. In X $A$ is $-3^\mathrm{h}20^\mathrm{m}$, that is $\mathrm{\Lambda}$ converted with $\mathrm{\bar{\Lambda}_{A}}=\ang{60;30}$ to $A=-3^\mathrm{h}18^\mathrm{m}$ and  rounded to the nearest $\nicefrac{1}{12}$\,h. The systematic difference of $\mathrm{\Delta \Lambda_A} = 2$\,min becomes 5\,min here.

\begin{figure}
\begin{center}
\includegraphics[width=120mm]{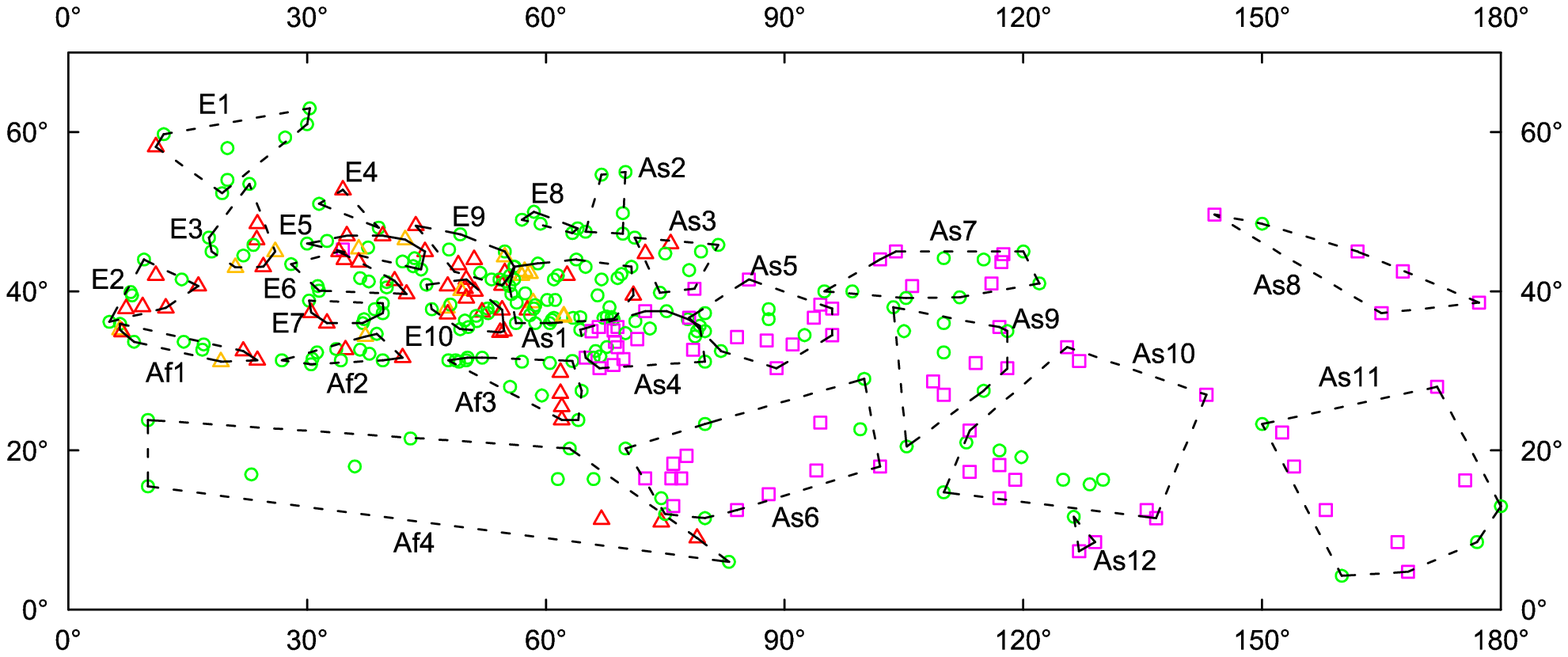}
\caption{Differences in the time differences $A_i$ to \textit{Alexandria} between Book 8 of the $\mathrm{\Omega}$-recension and the X-manuscript; \textit{green}/\textit{cirlce}: no difference in $A$; \textit{other:} difference in $A$; \textit{red}/\textit{triangle}: no difference in the longitude $\mathrm{\Lambda}$ of the location catalogue, \textit{yellow}/\textit{triangle}: difference in $\mathrm{\Lambda}$, \textit{magenta}/\textit{square}: no comparison of $\mathrm{\Lambda}$}
\label{fig:orte_dt_ox}
\end{center}
\end{figure}

Figure \ref{fig:orte_dt_ox} shows the localities without (green/circle) and with differences in $A$ between $\mathrm{\Omega}$ and X (Ptolemaic system). Those with differences are divided into localities with difference in $\mathrm{\Lambda}$ (yellow/triangle), without difference in $\mathrm{\Lambda}$ (red/triangle), and the remaining localities (magenta/square) which have no  $\mathrm{\Lambda}$ in the location catalogue of X. Localities with differences disperse across the whole area; however, they often occur in local groups. In the west, there is a larger amount of differences.

\section{Length of the longest day} \label{sec:laengster_tag}

The length of the longest day $M$ increases from $12$\,h at the equator ($\mathrm{\Phi}=0$) to $24$\,h for $\mathrm{\Phi}=90\arcdeg-\varepsilon$.
Ptolemy derives the computation of  $\mathrm{\Phi}$ from $M$ and vice versa in his \textit{Mathematike Syntaxis} (MS) 2.3. The modern formulation of that relation between  $M$ and $\mathrm{\Phi}$ is
\begin{equation} \label{eqn:M2B}
\cos{\left(  \frac{1}{2} M  \; 15 \nicefrac{\arcdeg}{\mathrm{h}} \right)} = - \tan{\mathrm{\Phi}} \; {\tan{\varepsilon}}
\end{equation}
($M$ in hours) with the obliquity of the ecliptic $\varepsilon$\footnote{Trigonometric problems could be solved accurately in antiquity by using tabulations of chords as it can be found in MS 1.11. On ancient trigonometry used by Ptolemy and described in his MS see \citet[p. 21\,ff.]{neu75}.}.
In MS~2.6 Ptolemy gives a compilation of parallels with specific $M$
including their latitudes (see Sect.~\ref{sec:laengster_tag:tab}).

Ptolemy does not explicitly mention  the $\varepsilon$ underlying the conversion between $\mathrm{\Phi}$ and $M$ in the Geography.
In GH~7.6.7  he gives the ratio  $\varepsilon:\nicefrac{\pi}{2}\approx\nicefrac{4}{15}$, that is the roughly rounded value $90\arcdeg \cdot \nicefrac{4}{15} = 24\arcdeg=:\varepsilon_\mathrm{r}$ for $\varepsilon$.
In MS~1.12 Ptolemy states that the arc between the solstice points (i.e. $2\varepsilon$) is approximately $(11/83)\cdot360\arcdeg$. That is $\ang{47;42;40}$ so that $\varepsilon$ becomes $\ang{23;51;20}=:\varepsilon_\mathrm{m}$ \citep[p. 44]{man12}.
Other appearances of $\varepsilon_\mathrm{m}$ are MS~1.14 and 2.4, where Ptolemy uses this value for exemplary computations, as well as the ``Table of Inclination'' in MS~1.15, where $\varepsilon_\mathrm{m}$ occurs as the size of the arc of meridian with an ecliptic longitude of $90\arcdeg$ (see Sect.~\ref{sec:dss}).
The correct value of Ptolemy's time is $\ang{23;41}=:\varepsilon_\mathrm{c}$ so that $\varepsilon_\mathrm{m}$  has an error of about $11\arcmin$.

\citet{raw85} gives two groups of cities of Book 8, on the one hand  10 cities, whose $M$ and $\mathrm{\Phi}$ could be related by Eq.~(\ref{eqn:M2B}) through $\ang{23;50}\approx\varepsilon_\mathrm{m}$, on the other hand 7 cities, whose $M$ and $\mathrm{\Phi}$ could be related  through $\ang{23;55}=:\varepsilon_\mathrm{h}$.
\citeauthor{raw85} attributes the latter value to Hipparchus \citep[see also][]{raw82} so that the original conversion would have been done with that value before Ptolemy.
A further hint to $\varepsilon_\mathrm{r}=24\arcdeg$ used in Ptolemy's data was detected by  \citet[pp. 236, 245\,ff.]{neu75}, who found  $\varepsilon_\mathrm{r}$ underlying the data in the tables for the first and last visibility of the planets in MS~13.10.
The question of the $\varepsilon$ underlying the $M_i$ in Book 8 is considered in the following.

\subsection{Conversion of coordinates and accuracy} \label{sec:umformung_M}

\begin{figure}
\centering
\subfigure{
\includegraphics[width=5.6cm]{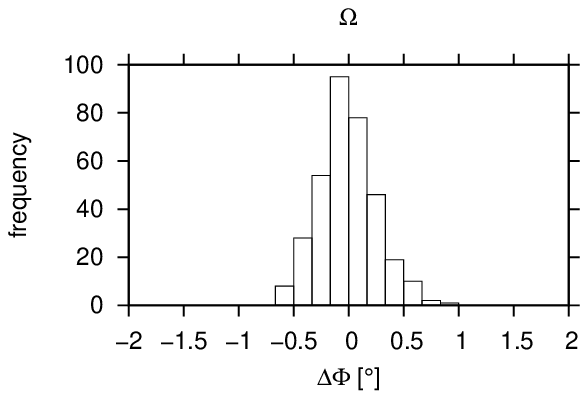}
}
\subfigure{
\includegraphics[width=5.6cm]{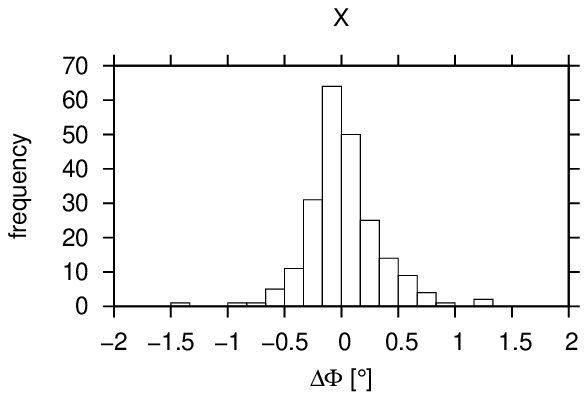}
}
\caption{Frequency of differences $\mathrm{\Delta \Phi}_i=\mathrm{\Phi}_i-\mathrm{\Phi}(M_i)$ between the latitudes $\mathrm{\Phi}_i$ of the location catalogue and the latitudes $\mathrm{\Phi}(M_i)$ computed from the lengths of the longest day $M_i$ of Book 8 by Eq.~(\ref{eqn:M2B}) for the $\mathrm{\Omega}$-recension (\textit{left}) and the X-manuscript (\textit{right})}
\label{fig:diff_b}
\end{figure}

A conversion between $\mathrm{\Phi}$ and $M$ was carried out for both directions $M\mapsto\mathrm{\Phi}$ and $\mathrm{\Phi}\mapsto M$ by means of Eq.~(\ref{eqn:M2B}), and the differences $\mathrm{\Delta\Phi}_i=\mathrm{\Phi}_i-\mathrm{\Phi}(M_i)$ and $\mathrm{\Delta} M_i=M_i-M(\mathrm{\Phi}_i)$ to the given values $\mathrm{\Phi}_i$ and $M_i$ were computed. For $M\mapsto\mathrm{\Phi}$ with $\varepsilon=\varepsilon_\mathrm{m}$ Fig.~ \ref{fig:diff_b} shows the frequencies of differences $\mathrm{\Delta \Phi}_i$ lying  within specified intervals (interval width $10\arcmin$). The distribution has a bell-shaped form with its maximum at $\approx0$. The majority of the absolute values of the differences is not larger than $\nicefrac{1}{2}\arcdeg$.

As a measure of the consistency between given coordinates $y_i$ and converted coordinates $y(x_i)$ the median absolute deviation
\begin{equation}
\label{eqn:mad}
\mathrm{MAD} = \mathrm{med}({|y_i - y(x_i)|})
\end{equation}
($i=1\ldots n$) and the mean deviation
\begin{equation}
\label{eqn:md}
\mathrm{MD} = \frac{1}{n}\sum_{i=1}^n{|y_i - y(x_i)|}
\end{equation}
are used here \citep[][p. 337]{ham74,sac92}. Unlike the MD, the MAD is resistent against outliers (i.e. abnormally large differences $y_i - y(x_i)$ here).

For both directions of conversion four conversions were carried out with $\varepsilon_\mathrm{m}$, $\varepsilon_\mathrm{h}$, $\varepsilon_\mathrm{r}$, as well as experimentally with $\varepsilon_\mathrm{c}$.
Table~\ref{tab:av_diff} shows the MAD, MD, and the number of coordinates $n$. Because of its resistent property the MAD is lower than the MD.
In most cases value $\varepsilon_\mathrm{m}$ leads to the lowest measures of precision; only in the case of $M\mapsto\mathrm{\Phi}$ the MAD based on $\varepsilon_\mathrm{h}$ is $1\arcmin$ (X) and $2\arcmin$ ($\mathrm{\Omega}$) smaller than the MAD based on $\varepsilon_\mathrm{m}$.
Regarding $\mathrm{\Phi}\mapsto M$ as the main direction, the results does not argue against $\varepsilon_\mathrm{m}$ underlying the data.

A conversion of coordinates and a comparison with given coordinate values was also carried out for the $M_i$ of Book 8 and the latitudes of the Canon.
For $\varepsilon=\varepsilon_\mathrm{m}$ the resulting MAD  slightly increases in comparison to the location catalogue  by $1\arcmin$ or $0.1$\,min, respectively,  and the MD increases  noticeably by $6\arcmin$ or $0.4$\,min, respectively (see Table~\ref{tab:av_diff}). Consequently, the close relation between the $M$-data of X and the latitudes of the Canon stated by  \citet[p. 44]{stu06} cannot be supported by means of the present computations.

\begin{figure}
\begin{center}
\includegraphics[width=120mm]{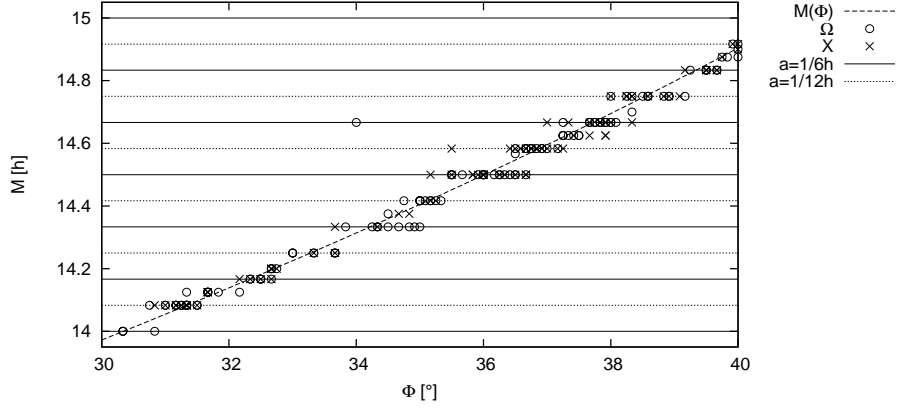}
\caption{Plot of the lengths of the longest day $M_i$ of Book 8 against the latitudes $\mathrm{\Phi}_i$ of the location catalogue for the $\mathrm{\Omega}$-recension and the X-manuscript, trigonometric conversion function $M(\mathrm{\Phi})$ (derived from Eq.~(\ref{eqn:M2B})); the horizontal lines show the position of $M$-values with resolutions of $a=\nicefrac{1}{6}$\,h and $a=\nicefrac{1}{12}$\,h}
\label{fig:M(B)_bsp}
\end{center}
\end{figure}

The size of the differences $\mathrm{\Delta\Phi}_i=\mathrm{\Phi}_i-\mathrm{\Phi}(M_i)$ can be explained by a rough resolution of the $M$-data and a disadvantageous propagation of their errors (assuming that the  original direction of conversion was $\mathrm{\Phi}\mapsto M$).
\citet[p. 104]{cun23} assumes  that the original resolution was $\nicefrac{1}{12}$\,h.
The largest denominators occurring in $\mathrm{\Omega}$ among the fractions of the $M_i$ are $150$ (\textit{Palura} (8.26.6)) and 60 (e.g. \textit{Ptolemais Herme} (8.15.13), \textit{Omanon} (8.22.12), \textit{Saue} (8.22.15));  however, they are exceptional  values. In X the largest denominator is $30$. In both $\mathrm{\Omega}$ and X the denominator $12$ occurs often so that the main resolution of $M$-values seems to be  $\nicefrac{1}{12}\,\mathrm{h} = 5\,\mathrm{min}$.
From fractions like $n/6=2n/12$ and $n/4=3n/12$ the precision is not known for sure, rougher resolutions than \nicefrac{1}{12}\,h are also possible. For $30\arcdeg\leq \mathrm{\Phi} \leq 40\arcdeg$ Fig.~\ref{fig:M(B)_bsp} exemplarily shows  the data points $(\mathrm{\Phi}_i, M_i)$ of $\mathrm{\Omega}$ and X as well as the curve of the trigonometric conversion function $M(\mathrm{\Phi})$ (derived from Eq.~(\ref{eqn:M2B}), $\varepsilon = \varepsilon_\mathrm{m}$). The data points are arranged in  horizontal groups, within which the $M$-values of different $\mathrm{\Phi}_i$ are equal. Most of these groups have a distance of $\nicefrac{1}{12}$\,h so that their resolution is apparently \nicefrac{1}{12}\,h (doted lines). Also $M_i$ with other resolutions become apparent. The plot shows, too, that equal $\mathrm{\Phi}_i$ were converted to different $M_i$. That can be explained by a rougher precision, i.e. $a=\nicefrac{1}{6}$\,h. Regarding only the lines of $a=\nicefrac{1}{6}$\,h (solid), no $\mathrm{\Phi}_i$ with different $M_i$ occur.

The error of an $M$ affects the corresponding latitude $\mathrm{\Phi}(M)$ according to the value of that latitude.
In general, the estimated effect of an error $\mathrm{\Delta}_x$ in $x$ on the value of the function $y(x)$ is
\begin{equation} \label{eqn:ff}
\mathrm{\Delta}_y = \frac{\mathrm{d}y}{\mathrm{d}x} \mathrm{\Delta}_x
\end{equation}
with the derivative $\mathrm{d}y/\mathrm{d}x$ of  $y(x)$ with respect to $x$. Applied to Eq.~(\ref{eqn:M2B}) the variation $\mathrm{\Delta}_{\mathrm{\Phi}}$ of $\mathrm{\Phi}$ according to the error $\mathrm{\Delta}_M$ of $M$ can be determined.

\begin{figure}
\centering
\subfigure{
\includegraphics[width=5.6cm]{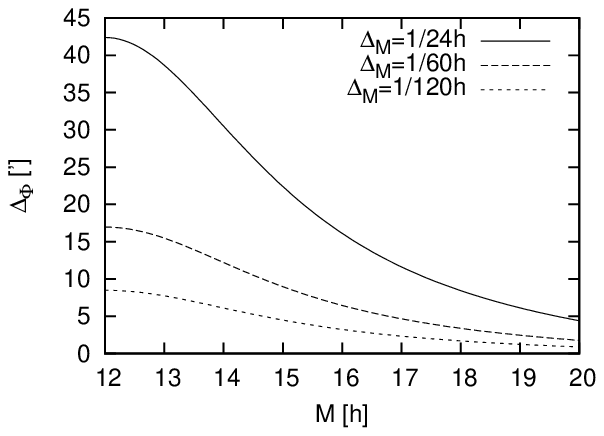}
}
\subfigure{
\includegraphics[width=5.6cm]{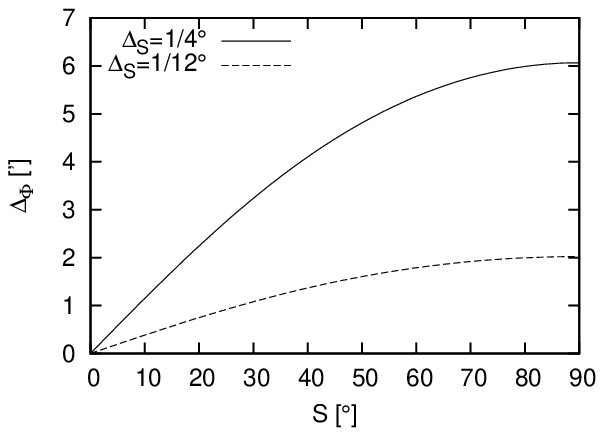}
}
\caption{Propagation of errors for the computation of the latitude $\mathrm{\Phi}$ from the length of the longest day $M$ (\textit{left}) and the distance $S$ from the summer solstice (\textit{right}) for different errors $\mathrm{\Delta}_M$ and $\mathrm{\Delta}_S$}
\label{fig:ff}
\end{figure}

Figure \ref{fig:ff} shows the propagation of errors for the range of $M$ in Book 8 ($12\,\mathrm{h}\leq M\leq 20\,\mathrm{h}$) and for the resolutions $a=$ \nicefrac{1}{60}\,h, \nicefrac{1}{30}\,h, and \nicefrac{1}{12}\,h; the variations were set to $\mathrm{\Delta}_M=a/2$ (Eq.~(\ref{eqn:rundung})).
Most of the localities in Book 8 have an $M<17\;\mathrm{h}$. For them a resolution of \nicefrac{1}{12}\,h can cause a significant loss of accuracy, the variation $\mathrm{\Delta}_{\mathrm{\Phi}}$ of $\mathrm{\Phi}$ is $>10\arcmin$. $\mathrm{\Delta}_{\mathrm{\Phi}}$ increases for $M\rightarrow 12$\,h to $\nicefrac{2}{3}\arcdeg$. Accordingly, an ancient conversion  $\mathrm{\Phi}\mapsto M$ with a precision of \nicefrac{1}{12}\,h (or more imprecise) caused a significant loss of accuracy in the latitude expressed by $M$. Even exceptionally high precisions as $\nicefrac{1}{60}$\,h could have a noticeable effect as the example of \textit{Ptolemais Herme} shows: $M=13\,\nicefrac{43}{60}$\,h, $\mathrm{\Delta}_M = 0.5\cdot\nicefrac{1}{60}$\,h, $\mathrm{\Delta_\Phi} = 7\arcmin$. Therefore, a large denominator in the fraction of $M$ does not generally mean that the value is given with an unrealistically high precision.

\subsection{Distribution of the coordinate differences} \label{sec:anpass_M}

For the coordinate differences $\mathrm{\Delta} M_i=M_i-M(\mathrm{\Phi}_i)$ after the conversion $\mathrm{\Phi}\mapsto M$ with $\varepsilon=\varepsilon_\mathrm{m}$ it was tested by means of the  $\chi^2$-test of goodness of fit \citep[e.g.][p. 420\,ff.]{sac92} whether they are normally distributed.  The test statistic is in general
\begin{equation}
\chi^2 = \sum_{i=1}^{k}{(n_i - n\,p_i)^2/(n\,p_i)}
\end{equation}
with the class number $k$, the observed frequencies $n_i$ of class $i$, the hypothetical probabilities $p_i$, and $n = \sum_{i=1}^{k}{n_i}$.
Null hypothesis of the test is the normal distribution assumption. Under this hypothesis $\chi^2$ follows  a $\chi^2$-distribution with $f = k - 1- u$ degrees of freedom, where  $u$ is the number of the unknown parameters. The two parameters of the normal distribution, the location parameter $\mu$ and the scale parameter $\sigma$, were estimated by the arithmetical mean of the $\mathrm{\Delta} M_i$ and by the root mean square deviation (with respect to the mean).
A few outlying, very large differences, which are not characteristic for the distribution form, were excluded. According to the assumed crudest resolution $a=\nicefrac{1}{4}$\,h of the $M_i$ (see Sect.~\ref{sec:schiefe}) the limit for $|\mathrm{\Delta} M_i|$ was set to $a/2=7.5$\,min (Eq.~(\ref{eqn:rundung})). The level of significance $\alpha$ used is 5\%.
The results are:
\begin{compactitem} [--]
\item $\mathrm{\Omega}$: $n=341$, $k=10$, $f=7$, $\mu=0.06$\,min, $\sigma=1.58$\,min, $\chi^2= 26.9$;
\item X: $n=216$, $k=10$, $f=7$, $\mu=-0.17$\,min, $\sigma=1.83$\,min, $\chi^2= 27.5$.
\end{compactitem}
The quantile is in both cases $q_{1-\alpha,f}=14.1$ so that the normal distribution assumption has to be rejected.

The $\chi^2$-test was repeated for the Laplace distribution. Its two parameters $\mu$ and $\sigma$ were estimated by the median and by the MD (Eq.~(\ref{eqn:md}), but here with respect to the computed median). The results are:
\begin{compactitem} [--]
\item $\mathrm{\Omega}$: $n=341$, $k=10$, $f=7$, $\mu=0.06$\,min, $\sigma=1.16$\,min, $\chi^2= 12.4$;
\item X: $n=216$, $k=10$, $f=7$, $\mu=0.03$\,min, $\sigma=1.35$\,min, $\chi^2= 7.6$.
\end{compactitem}
For $\mathrm{\Omega}$ and X $\chi^2$ goes below the quantile $q_{1-\alpha,f}=14.1$. Accordingly it can be assumed that  the differences $\mathrm{\Delta}M_i$ follow (at least approximately) a Laplace distribution.

\subsection{On the original direction of conversion} \label{sec:richtung_M}

Ptolemy states in GH~8.2.1 that the $M$-data are computed from latitudes. However, it is known from a few localities that Ptolemy adopted their $M$-values from Hipparchus  \citep[p. 29]{ber00}. Hipparchus attributed them to parallels with particular lengths of the longest day (\textit{Mero\"{e}} (8.16.9), \textit{Ptolemais Theron} (8.16.10), \textit{Soene} (8.15.15), \textit{Rhodos} (8.17.21), and \textit{Massalia} (8.5.7)). From these localities can be expected that their latitudes in the location catalogue are derived from $M$ in Book 8.
\citet{raw85,raw08} mentions the possibility of a conversion $M \mapsto \mathrm{\Phi}$ for major cities; in \citet{raw85} examples are given.
Exemplarily, the localities shall be considered which \citeauthor{raw85} relates to $\varepsilon_\mathrm{h}$.
The contrary conversion $\mathrm{\Phi}\mapsto M$ was done by means of Eq.~(\ref{eqn:M2B}) with $\varepsilon_\mathrm{m}$; the result is shown in Table~\ref{tab:raw}.
As can be seen, each computed $M(\mathrm{\Phi})$ agrees satisfactorily with the value in Book 8. Hence, the coordinates of these localities also support the direction of conversion $\mathrm{\Phi} \mapsto M$.

To provide evidence for the original direction of conversion it was examined how satisfactorily  the coordinate differences $\mathrm{\Delta\Phi}_i=\mathrm{\Phi}_i - \mathrm{\Phi}(M_i)$ and $\mathrm{\Delta}M_i= M_i - M(\mathrm{\Phi}_i)$ after both conversions $M\mapsto\mathrm{\Phi}$ and $\mathrm{\Phi}\mapsto M$ can be explained by the precision of the coordinates.
Unfortunately, the precision of $\mathrm{\Phi}$ and $M$ is not constant and not known in each case; however, a maximal resolution can be given for the $\mathrm{\Phi}$-values and a main resolution for the $M$-values. Thereby limits $L_\mathrm{\Delta\Phi}$ and $L_{\mathrm{\Delta}M}$ for the differences $\mathrm{\Delta\Phi}_i$ and $\mathrm{\Delta}M_i$ can be given so that the numbers of differences was determined, which go below these limits.

For the  $\mathrm{\Phi}$-values of the \textit{poleis episemoi} a maximal resolution of $\nicefrac{1}{2}\arcdeg$ can be assumed, definitely the actual resolution is often more precise (e.g. $\nicefrac{1}{12}\arcdeg$ or $\nicefrac{1}{6}\arcdeg$,  \citep[see][]{mar11}). According to Eq.~(\ref{eqn:rundung}) the limit $L_\mathrm{\Delta\Phi}$ is then $0.5\cdot\nicefrac{1}{2}\arcdeg=\nicefrac{1}{4}\arcdeg$. For the $M$-values a main resolution of \nicefrac{1}{12}\,h  (or more imprecise; see Sect.~\ref{sec:umformung_M}) can be assumed so that  $L_{\mathrm{\Delta}M} = 0.5 \cdot \nicefrac{1}{12}\,\mathrm{h}= \nicefrac{1}{24}\,\mathrm{h}=2.5\,\mathrm{min}$ was used as a limit.

For both conversions $M \mapsto \mathrm{\Phi}$ and $\mathrm{\Phi}\mapsto M$ the parameters  $\varepsilon_\mathrm{m}$ and $\varepsilon_\mathrm{h}$ were used. The result is:
\begin{compactitem} [--]
\item $\varepsilon_\mathrm{m}$: $\mathrm{\Omega}$: $69\%$ $|\mathrm{\Delta\Phi}| \leq L_\mathrm{\Delta\Phi}$, $88\%$ $|\mathrm{\Delta}M| \leq L_{\mathrm{\Delta}M}$;
    X: $66\%$ $|\mathrm{\Delta\Phi}| \leq L_\mathrm{\Delta\Phi}$, $82\%$ $|\mathrm{\Delta}M| \leq L_{\mathrm{\Delta}M}$;
\item $\varepsilon_\mathrm{h}$: $\mathrm{\Omega}$: $68\%$ $|\mathrm{\Delta\Phi}| \leq L_\mathrm{\Delta\Phi}$, $87\%$ $|\mathrm{\Delta}M| \leq L_{\mathrm{\Delta}M}$;
    X: $66\%$ $|\mathrm{\Delta\Phi}| \leq L_\mathrm{\Delta\Phi}$, $81\%$ $|\mathrm{\Delta}M| \leq L_{\mathrm{\Delta}M}$.
\end{compactitem}
The test conditions for the conversions $M \mapsto \mathrm{\Phi}$ and $\mathrm{\Phi}\mapsto M$ were not the same, for $\mathrm{\Delta\Phi}$ the limit was advantageously large and for $\mathrm{\Delta}M$ small. Nevertheless, the amount of $\mathrm{\Delta}M_i$ going below their limit is noticeably larger than in the case of $\mathrm{\Delta\Phi}_i$. Thus, the results argue for a main direction of conversion of $\mathrm{\Phi}\mapsto M$.
The resulting $M_i$ of this conversion were round values, whose inaccuracies affect the $\mathrm{\Delta\Phi}_i$ computed here after the back conversion $M\mapsto \mathrm{\Phi}$.

\subsection{The obliquity of the ecliptic} \label{sec:schiefe}

The question of which value $\varepsilon$ underlies the $M$-data was investigated more in detail chapter by chapter, assuming that the majority of the localities of a chapter were converted with the same value of $\varepsilon$.
An approach to determine the underlying $\varepsilon$ is an adjustment of $\varepsilon$, which also leads to information about the accuracy of the result.
Solving Eq.~(\ref{eqn:M2B}) for $M$ gives the observation equation of the adjustment model
\begin{equation} \label{eqn:beogl_M}
M_i + v_i = f(\varepsilon) = 2  \frac{1\,\mathrm{h}}{15\arcdeg} \, \arccos{\left(  - \tan{\mathrm{\Phi}_i} \; {\tan{\varepsilon}}\right)} \; , \quad i = 1,\ldots, n
\end{equation}
with the observation $M_i$ and its correction $v_i$.
The solution of this non-linear adjustment problem can be obtained iteratively by means of a linearisation \citep[see e.g.][p. 116]{jae05}.
In Sect.~\ref{sec:anpass_M} the $\chi^2$-test of goodness of fit  showed that the differences $\mathrm{\Delta} M_i$ with respect to a conversion with $\varepsilon_\mathrm{m}$  follow rather a Laplace distribution than a normal distribution. It shall be assumed here that the Laplace distribution can also be expected for errors of $M_i$ whose conversion is based on $\varepsilon \neq \varepsilon_\mathrm{m}$. The maximum likelihood estimation for the Laplace distribution is the $L_1$-norm estimation, which is applied here \citep[see e.g.][p. 117\,ff.]{jae05}.

To estimate the accuracy of the adjusted parameter $\varepsilon$, the term $q=(\mathbf{A}^\top \mathbf{A})^{-1}$ is needed, whose one-column design matrix $\mathbf{A}$ of the linearised adjustment problem contains the derivatives $\mathrm{d}f_i(\varepsilon)/\mathrm{d}\varepsilon$ of $f_i$ with respect to $\varepsilon$.
Then the $(1-\alpha)100\%$ confidence interval for $\varepsilon$ can be estimated according to \citet{die82, die90} by
\begin{equation} \label{eqn:eps_genauigkeit}
\varepsilon \pm \mathrm{\Delta}\varepsilon = \varepsilon \pm z_{\alpha/2}\lambda \sqrt{q}
\end{equation}
with the quantile $z_{\alpha/2}$ of the standard normal distribution as well as with
\begin{equation}
\lambda = \frac{\hat{v}_{m+\mu} - \hat{v}_{m-\nu}}{4\nu/n} \; .
\end{equation}
Value $m$ is the index of the median of the computed corrections $v_i$ in the ordered sequence $\hat{v}_i$. Values $\hat{v}_{m+\mu}$ and $\hat{v}_{m-\nu}$ are the corrections having a distance $\nu$ from $\hat{v}_m$ in the ordered sequence.

The investigations in Sects.~\ref{sec:umformung_M} and \ref{sec:richtung_M} show that there are inaccurate $M_i$ with larger errors, possibly partly corrupt\footnote{Examples for corrupt or roughly given $M$-values are \textit{Nisaia} (8.23.6) and \text{Arbis} (8.25.14) in $\mathrm{\Omega}$ \citep[p. 205, 209]{stu09}}.
The $L_1$-norm estimation of $\varepsilon$ and the estimation (\ref{eqn:eps_genauigkeit}) of $\mathrm{\Delta}\varepsilon$ have the advantage to be resistent against outlying observations.

Due to their small number of observations ($\leq3$) the chapters Europe~4 and Asia~12 were excluded from the parameter estimation.
$\lambda$ was computed using all $v_i$ of the remaining chapters.
$\nu$ must to be small but larger than the half of the number $u$ of adjustments of $\varepsilon$; it was set to $11+u/2=18$ for X and to $18+u/2=30$ for $\mathrm{\Omega}$\footnote{In an L$_1$-norm estimation of $u$ unknown parameters, (at least) $u$ corrections become zero. The choice $\nu'=11=\nu-u/2$ for X is reconcilable with the values used in \citet{die90}. According to the ratio $(n_\mathrm{\Omega}-u_\mathrm{\Omega})/(n_X-u_X)=1.6$, $\nu$ was set to $\nu=1.6\nu'+u/2=18+12=30$ for $\mathrm{\Omega}$.}.
For the confidence interval of $\varepsilon$  the probability 95\% was chosen. In summary, the results are
\begin{compactitem} [--]
\item $\mathrm{\Omega}$: $n=338$, $u=24$, $\lambda=0.7\arcmin$, smallest $\varepsilon$: $\ang{23;46}\pm 8\arcmin$ in Europe~10, largest $\varepsilon$: $\ang{24;07}\pm 30\arcmin$ in Asia~11;
\item X:  $n=217$, $u=16$, $\lambda=1.0\arcmin$, smallest $\varepsilon$: $\ang{23;30}\pm 44\arcmin$ in Africa~4 (anomalously inaccurate) and  $\ang{23;45}\pm 11\arcmin$ in  Europe~10, largest $\varepsilon$: $\ang{23;56}\pm 11\arcmin$ in Europe~9.
\end{compactitem}
With the exception of Asia~11 in $\mathrm{\Omega}$ and Africa~4 in X the adjusted values of $\varepsilon$ deviate from $\varepsilon_\mathrm{m}$ in the order of some arc minutes ($\leq10\arcmin$). Only in the case of Asia~3 in $\mathrm{\Omega}$  $\varepsilon_\mathrm{m}$ lies  outside of the estimated $95\%$-confidence interval ($\varepsilon = \ang{24;02}\pm 9\arcmin$). Apart from this exception, the adjustment does not result in significant deviations from $\varepsilon_\mathrm{m}$ and accordingly the results do not argue against $\varepsilon_\mathrm{m}$ underlying the $M$-values in general.

Moreover, $\varepsilon_\mathrm{m}$ need not be rejected as long as the differences $\mathrm{\Delta}M_i=M_i-M(\mathrm{\Phi}_i)$ based on $\varepsilon_\mathrm{m}$ can be explained by rounding.
Exemplarily Asia~3 in $\mathrm{\Omega}$ with the significant deviation from $\varepsilon_\mathrm{m}$ is considered. Each of its ten $M_i$ can be explained by  a kind of rounding as it can be found among the $M_i$ of Book 8 (see Sect.~\ref{sec:laengster_tag:tab}). There are eight $M_i$ explicable by a rounding to the nearest $\nicefrac{1}{12}$\,h. Further, in the case of \textit{Thospia} the precision could be \nicefrac{1}{24}\,h ($M(\mathrm{\Phi}=\ang{39;50})=14^\mathrm{h}53.2^\mathrm{m}\approx14\,\nicefrac{21}{24}\,\mathrm{h}=14\,\nicefrac{7}{8}\,\mathrm{h} =M$) and in the case of \textit{Albana} \nicefrac{1}{6}\,h ($M(\mathrm{\Phi}=\ang{45;50})=15^\mathrm{h}36.7^\mathrm{m}\approx15\,\nicefrac{4}{6}\,\mathrm{h}=14\,\nicefrac{2}{3}\,\mathrm{h} =M$).
Apart from that, $M$ of \textit{Artemita} is not explicable by a conversion with $\varepsilon_\mathrm{r}$ (which is to be expected in view of the adjustment) in conjunction with any kind of rounding ($M(\mathrm{\Phi}=\ang{40;20})=14^\mathrm{h}57.7^\mathrm{m}$, but $M = 14^\mathrm{h}55.0^\mathrm{m}$).

Finally, all $M_i$ were used for an L$_1$-norm-adjustment of one parameter $\varepsilon$. For $\mathrm{\Omega}$ is obtained $\varepsilon = \ang{23;52}\pm 2\arcmin$ and for X $\varepsilon = \ang{23;52}\pm 3\arcmin$.
Among the different values for $\varepsilon$ introduced here, only $\varepsilon_\mathrm{m}$ lies within both 95\% confidence intervals.

\subsection{Conversion by means of linear interpolation} \label{sec:laengster_tag:tab}

It is much more probable that a conversion of coordinates was originally done by means of a tabulation and not by means of a single, trigonometric computation from Eq.~(\ref{eqn:M2B}) for each locality.
\citet[p. 101\,ff.]{cun23} assumes an underlying tabulation $(M_j, \mathrm{\Phi}(M_j))$ with a step size of $\nicefrac{1}{12}\,\mathrm{h}=5\,\mathrm{min}$. Using this tabulation, \citeauthor{cun23} cannot explain each $M$-value in the examined chapters (8.3--8). An example is \textit{Gesoriacum} (8.5.6) with $\mathrm{\Phi}=\ang{53;30}$ and $M=16^\mathrm{h}50^\mathrm{m}$. The nearest $\mathrm{\Phi}$ in \citeauthor{cun23}'s tabulation is $\ang{53;39}$ which leads to $M=16^\mathrm{h}55^\mathrm{m}$. Thus, \citeauthor{cun23} supposes that $\mathrm{\Phi}=\ang{53;30}$ is not the original value underlying $M$.

Another approach to explain the $M$-data in Book 8 is taking recourse to known $\mathrm{\Phi}$- and $M$-data given by Ptolemy.
There are two datasets with latitudes and corresponding $M$-data in the Geography and the MS, these are the compilations of parallels in GH~1.23 and MS~2.6. It is conceivable that such data were used in conjunction with an interpolation, a method applied by Ptolemy \citep[p. 84]{ped11}.
Considering the mentioned example of \textit{Gesoriacum}, $M(\mathrm{\Phi})$ computed from Eq.~(\ref{eqn:M2B}) is $16^\mathrm{h}54^\mathrm{m}$. The same result is obtained by means of the compilation of the parallels in GH~1.23 and a linear interpolation.  The slope $a$ of the 17th and 18th parallels in GH~1.23 is
\begin{equation}
\begin{split}
a &= (M_{j+1}-M_j)/(\mathrm{\Phi}_{j+1} - \mathrm{\Phi}_j) \\
&= (17^\mathrm{h}00^\mathrm{m}-16^\mathrm{h}30^\mathrm{m})/(\ang{54;00}-\ang{51;30} ) = 0.2\,  \nicefrac{\mathrm{h}}{\arcdeg} \; .
\end{split}
\end{equation}
The approximated $M$ of \textit{Gesoriacum} is then
\begin{equation}
M = 16^\mathrm{h}30^\mathrm{m} + a \cdot (\mathrm{\Phi} - \ang{51;30}) = 16^\mathrm{h}54^\mathrm{m} \; .
\end{equation}
This value rounded to the nearest $\nicefrac{1}{6}$ is $16^\mathrm{h}50^\mathrm{m}$, the value in the location catalogue. As the plot in Fig.~\ref{fig:M(B)_bsp} shows, a precision of $\nicefrac{1}{6}$\,h can be assumed for part of the $M$-data (see Sect.~\ref{sec:umformung_M}).

\begin{figure}
\begin{center}
\includegraphics[width=120mm]{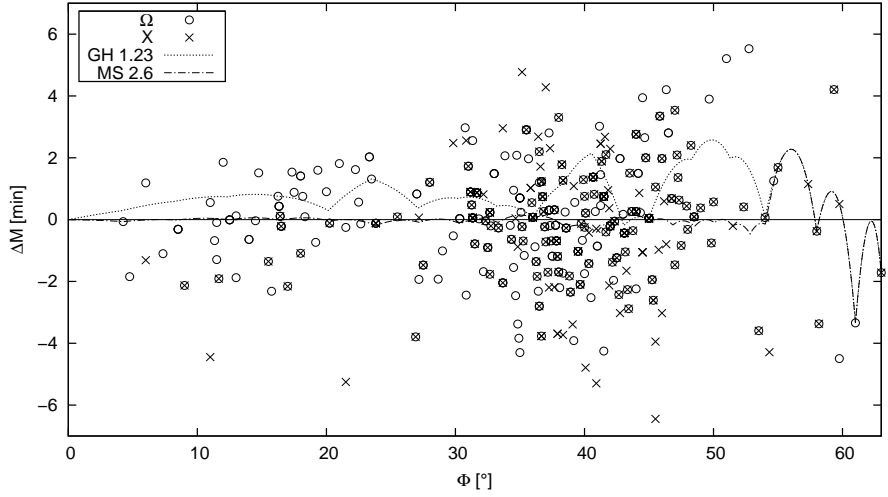}
\caption{Differences $\mathrm{\Delta}M=M-M(\mathrm{\Phi})$ between the length of the longest day  $M(\mathrm{\Phi})$, trigonometrically computed from the latitude $\mathrm{\Phi}$ (Eq.~(\ref{eqn:M2B})), and the $M$-data of Book 8 ($\mathrm{\Omega}$-recension and X-manuscript) as well as $M$ resulting from an linear interpolation of the $M$-data of GH~1.23 and MS~2.6}
\label{fig:diff_interpol}
\end{center}
\end{figure}

The accuracy of a linear interpolation of  the data in GH~1.23 and MS~2.6 was determined by computing the difference $\mathrm{\Delta}\tilde{M}(\mathrm{\Phi})=\tilde{M}(\mathrm{\Phi})-M(\mathrm{\Phi})$ between the interpolated value $\tilde{M}(\mathrm{\Phi})$ and the trigonometric computed value $M(\mathrm{\Phi})$ (Eq.~(\ref{eqn:M2B})). Figure \ref{fig:diff_interpol} shows $\mathrm{\Delta} \tilde{M}(\mathrm{\Phi})$ and additionally the differences $\mathrm{\Delta} M_i=M_i-M(\mathrm{\Phi_i})$ between the $M_i$ of Book 8 and the $M(\mathrm{\Phi}_i)$ computed from the $\mathrm{\Phi}_i$ of the location catalogue (a few data points with $|\Delta M|>7$\,min are not plotted). For $\mathrm{\Phi}>54\arcdeg$ the curves of the interpolated GH- and MS-data are equal\footnote{There are more parallels in MS~2.6 than in GH~1.23 from $45\arcdeg$ latitude, but from $54\arcdeg$ to $63\arcdeg$ there are some interpolated values in MS~2.6 so that a linear interpolation of the data in MS~2.6 and in GH~1.23 leads to the same result for latitudes  $>54\arcdeg$.}. For $\mathrm{\Phi}<54\arcdeg$ the interpolation of the MS-data is very accurate ($|\mathrm{\Delta} \tilde{M}| < 0.5$\,min); for $\mathrm{\Phi}\geq 54\arcdeg$ the majority of the differences $\mathrm{\Delta}M_i$ coincide with the curve $\mathrm{\Delta}\tilde{M}(\mathrm{\Phi})$. This indicates that the $M_i$ of Book 8 are based on an interpolation of the data in MS~2.6.
The interpolation of the GH-data produces errors up to 3\,min. Its curve $\mathrm{\Delta}\tilde{M}(\mathrm{\Phi})$ does not coincide with the distribution of the differences $\mathrm{\Delta}M_i$ in general. However, some data points lie next to the curve, and three data points coincide with the curve so that an interpolation of the data in GH~1.23 can be assumed at least in these cases ($\mathrm{\Omega}$: \textit{Harmuza} (8.22.21), \textit{Aspithra} (8.27.11); $\mathrm{\Omega}$ and X: \textit{Ambrodax} (8.21.17)).

Finally, the trigonometric computation ($\varepsilon=\varepsilon_\mathrm{m}$) and the linear interpolation were compared with regard to the amount of $M_i$ explicable by their precision.
For this purpose, Procedure \texttt{ROUND} was applied (see Sect.~\ref{sec:alex}).
As it was shown in Sect.~\ref{sec:umformung_M}, different resolutions come into consideration.
The possible resolutions of $M$ were chosen according to the denominators of the fractions occurring in Book 8. In addition to the standard resolution $a_\mathrm{s}=\nicefrac{1}{12}$\,h  there are in $\mathrm{\Omega}$ the more precise $a$ (in h) $\nicefrac{1}{15}$, $\nicefrac{1}{20}$, $\nicefrac{1}{24}$, $\nicefrac{1}{30}$, $\nicefrac{1}{60}$, and $\nicefrac{1}{150}$ as well as the less precise $\nicefrac{1}{10}$, $\nicefrac{1}{8}$, $\nicefrac{1}{6}$, and $\nicefrac{1}{4}$.
The more precise $a$ in X are (in h) $\nicefrac{1}{15}$ and $\nicefrac{1}{30}$ as well as the less precise $\nicefrac{1}{8}$, $\nicefrac{1}{6}$, and $\nicefrac{1}{4}$.
The result is given in Table~\ref{tab:aufloes_T}. The more precise the resulting resolutions are, the more probable the conversion can be considered. The trigonometric conversion and the interpolation of the MS-data have nearly the same result; the interpolation of the data in GH~1.23 has a worse result than the other two methods. Thus, a conversion by means of the data in MS~2.6 is very probable. The amount of $M_i$ being explainable with a resolution of \nicefrac{1}{12}\,h is 72\% for $\mathrm{\Omega}$ and 69\% for X.

The results in Table~\ref{tab:aufloes_T} show that in the case of $\mathrm{\Omega}$ 84\% and in the case of X 72\% of the localities have only small absolute differences $<0.5 \cdot \nicefrac{1}{12}\,\mathrm{h}=\nicefrac{1}{24}\,\mathrm{h}=2.5\,\mathrm{min}$ (cf. Eq.~(\ref{eqn:rundung})) between $M_i$ and $M(\mathrm{\Phi}_i)$ converted by means of the data of MS~2.6.
Almost all $M_i$ (ca. 95\%) can be explained by a conversion (data of MS~2.6 or Eq.~(\ref{eqn:M2B})) and rounding. Hence, the majority of the differences between Book 8 and the location catalogue are explicable by the precision of the $M$-data. Accordingly, both datasets show a high level of interdependence.

There is a lower percentage of $M_i$ (3\% in $\mathrm{\Omega}$, 6\% in X) whose differences $\mathrm{\Delta}M_i$ are inexplicable by one of the types of rounding used. These as well as lower precise explicable $M$-values occur in nearly all chapters of Book 8. Possible reasons for $M_i$ and $M(\mathrm{\Phi}_i)$ not fitting  are
\begin{compactenum} [(1)]
\item $M$ is rounded incorrectly\footnote{A known inconsistence  of this kind, but in latitude, occurs in GH~1.23 in the latitude $\mathrm{\Phi}=\ang{43;05}$ of the 13th parallel. Since $M$ of this parallel is $15^\mathrm{h}15^\mathrm{m}$, the correct value for $\mathrm{\Phi}$ should be $\ang{43;00}\approx \mathrm{\Phi}(M)=\ang{43;01}$ which was given by Ptolemy in MS~2.6.} or is slightly erroneous due to a propagation of rounding errors in intermediate results; for example possibly \textit{Asturica August} (8.4.5)  with $M=15^\mathrm{h}25^\mathrm{m}$ (in $\mathrm{\Omega}$ and X) and $M(\mathrm{\Phi})=15^\mathrm{h}22.2^\mathrm{m} \approx 15^\mathrm{h}20^\mathrm{m}$;
\item $M$ is wrong due to a corruption; for example possibly \textit{Palmyra} (8.20.10) in $\mathrm{\Omega}$  with $M=14^\mathrm{h}40^\mathrm{m}$ and $M(\mathrm{\Phi})=14^\mathrm{h}19^\mathrm{m} \approx 14^\mathrm{h}20^\mathrm{m}$;
\item $\mathrm{\Phi}$ is wrong due to a corruption; for example \textit{Osika} (8.19.9) in X with $\mathrm{\Phi}=\ang{45;30}$, $M=15^\mathrm{h}27.5^\mathrm{m}$, $M(\mathrm{\Phi})=15^\mathrm{h}34^\mathrm{m}$ (in contrast $\mathrm{\Omega}$: $\mathrm{\Phi}=\ang{44;45}$ and Canon: $\mathrm{\Phi}=\ang{44;30}$).
\end{compactenum}

It was tested for the remaining inexplicable $M_i$, whether a conversion with $\varepsilon_\mathrm{h}$ or $\varepsilon_\mathrm{r}$ brings an explanation of their fractions. For this purpose Procedure \texttt{ROUND} was repeated. As a result, a few $M_i$ can be explained by a rounding, but there still remain inexplicable values.
However, it is not necessary to assume another value than $\varepsilon_\mathrm{m}$;  each of the problematic $M_i$ can be explained by Error 1: \textit{Asturica August}: see above; \textit{Luppia} (8.6.3): $M(\mathrm{\Phi})=16^\mathrm{h}44^\mathrm{m}$ became $16^\mathrm{h}50^\mathrm{m}$ instead of $16^\mathrm{h}40^\mathrm{m}$ (resolution \nicefrac{1}{6}\,h); \textit{Ravenna} (8.8.5): as \textit{Asturica August}; \textit{Capua} (8.8.6): $M(\mathrm{\Phi})=15^\mathrm{h}02^\mathrm{m}$ became $15^\mathrm{h}05^\mathrm{m}$ instead of $15^\mathrm{h}00^\mathrm{m}$; \textit{Mallos} (8.17.44): $M(\mathrm{\Phi})=14^\mathrm{h}32.8^\mathrm{m}$ became $14^\mathrm{h}34^\mathrm{m}$ instead of $14^\mathrm{h}32^\mathrm{m}$ (resolution \nicefrac{1}{30}\,h); \textit{Medaba} (8.20.20): $M(\mathrm{\Phi})=14^\mathrm{h}02^\mathrm{m}$ became $14^\mathrm{h}05^\mathrm{m}$ instead of $14^\mathrm{h}00^\mathrm{m}$.

\section{Distance from the summer solstice} \label{sec:dss}

At places located between the tropics the sun is at the zenith two (for $|\mathrm{\Phi}|<\varepsilon$) or one (for $|\mathrm{\Phi}|=\varepsilon$) times within the year. For this date, the distance $S$ of the sun from the summer solstice on the ecliptic is given in Book 8.
The quantity $l=90\arcdeg-S$ is the ecliptic longitude, reckoned from the vernal equinox.
There are about 60 localities in Book 8 with given $S$.
In MS~1.14, Ptolemy explains the computation of the distance of the sun from the equator (declination) by means of $l$\footnote{In MS~1.14, the ecliptic longitude $l$ is reckoned from the autumnal equinox.}. When the sun is at the zenith, its declination equals the latitude $\mathrm{\Phi}$  of the locality. Thus, the calculational relations can be used to determine $l$ or $S$ from $\mathrm{\Phi}$.
The relation between
$S$ and the latitude $\mathrm{\Phi}$ is
\begin{equation} \label{eqn:S2B}
\sin{ \mathrm{\Phi} } = \sin{\varepsilon} \; \cos{S} \;.
\end{equation}

In MS~1.15, Ptolemy gives a list of tabulated $\mathrm{\Phi}(l_j)$ for $l_j=1\arcdeg,\ldots,90\arcdeg$ in steps of  $1\arcdeg$ (``Table of Inclination''). A further source of $S$-data is MS~2.6, where Ptolemy arranges data about specific parallels (see Sect.~\ref{sec:laengster_tag:tab}); however, it includes only 5 data pairs $(\mathrm{\Phi}_j, S_j)$. For the parallels with $\mathrm{\Phi}< \varepsilon$ $S$ is given there.
In the following  both compilations of $S$-data in the MS are taken into consideration besides Eq.~(\ref{eqn:S2B}) as a possible basis of the $S$-data in Book 8 of the Geography.

\subsection{Conversion of coordinates and accuracy}

Firstly, the precision of the $S_i$ is considered to get an idea, which size of coordinate differences are to be expected after a conversion between $S$ and $\mathrm{\Phi}$.
The largest denominator of the fractions of the $S_i$ is 6 in $\mathrm{\Omega}$ and 8 in X; the main resolutions seem to be $\nicefrac{1}{3}\arcdeg$ and $\nicefrac{1}{4}\arcdeg$.
Solving Eq.~(\ref{eqn:S2B}) for $\mathrm{\Phi}$, the variation $\mathrm{\Delta}_{\mathrm{\Phi}}$ of $\mathrm{\Phi}$ due to an error  $\mathrm{\Delta}_S$ in $S$ can be estimated by Eq.~(\ref{eqn:ff}).
The computation of propagation of errors was done exemplarily for the  resolutions $\nicefrac{1}{6}\arcdeg$ and $\nicefrac{1}{2}\arcdeg$. The variations were set to the expected maximal rounding errors (Eq.~(\ref{eqn:rundung})): $\mathrm{\Delta}_S=0.5\cdot\nicefrac{1}{6}\arcdeg=\nicefrac{1}{12}\arcdeg$ and $\mathrm{\Delta}_S=0.5\cdot\nicefrac{1}{2}\arcdeg= \nicefrac{1}{4}\arcdeg$.
Figure \ref{fig:ff} shows the result.
The effect of a variation in $S$ on $\mathrm{\Phi}$ is very low (in comparison to $M$ and $\mathrm{\Phi}$). For an inaccurate resolution of $\nicefrac{1}{2}\arcdeg$ the expected maximal error in $\mathrm{\Phi}$ increases only to $6\arcmin$; it goes to zero for $S\rightarrow 0$ (i.e. $\mathrm{\Phi} \rightarrow \varepsilon$).

The conversion between $S$ and $\mathrm{\Phi}$ was carried out for both directions $S\mapsto \mathrm{\Phi}$ and $\mathrm{\Phi}\mapsto S$ by means of Eq.~(\ref{eqn:S2B}) with the four different $\varepsilon$ introduced in Sect.~\ref{sec:laengster_tag}, and the differences $\mathrm{\Delta\Phi}_i=\mathrm{\Phi}_i-\mathrm{\Phi}(S_i)$ and $\mathrm{\Delta} S_i=S_i-S(\mathrm{\Phi}_i)$ to the given values $\mathrm{\Phi}_i$ and $S_i$ were computed.
Since X  ends in GH~5.13, only 16 $S$-values can be considered for the comparison of Book 8 and the location catalogue (these are \textit{Syene} (8.15.15) and Chapter 16 (Asia~4)).
From the differences between the converted and the original coordinates, the MAD and the MD were computed by Eqs.~(\ref{eqn:mad}) and (\ref{eqn:md}); the results are given in Table~ \ref{tab:av_diff}\footnote{In the case of $\mathrm{\Phi}\mapsto S(\varepsilon_\mathrm{c})$ the number $n$ of the converted latitudes is smaller because values $\mathrm{\Phi}>\varepsilon_\mathrm{c}$ cannot be converted by Eq.~(\ref{eqn:S2B}). \textit{Arbis} (8.25.14) is excluded from the computations because of its exceptionally wrong $S=0\arcdeg$ in $\mathrm{\Omega}$.}.

In the case of X, the $\mathrm{MAD}=2\arcmin$  for $\mathrm{\Phi}\mapsto S(\varepsilon_\mathrm{m})$ is very small and goes below the expected result  with regard to the assumed main precisions of $\nicefrac{1}{3}\arcdeg$ and $\nicefrac{1}{4}\arcdeg$.
In contrast, the $\mathrm{MD}=31\arcmin$ is considerably larger. It is increased by four large differences\footnote{These localities are \textit{Syene} (8.15.15) and \textit{Autolalai} (8.16.3) with the crude value $S=0\arcdeg$ for $\mathrm{\Phi}=\ang{23;50}$ as well as \textit{Dere} (8.16.12) and \textit{Mosylon} (8.16.13).}.
The MAD for $\varepsilon \neq \varepsilon_\mathrm{m}$ is significantly larger than for $\varepsilon_\mathrm{m}$; that argues for $\varepsilon_\mathrm{m}$ underlying the $S$-data of X.

\begin{figure}
\begin{center}
\includegraphics[width=120mm]{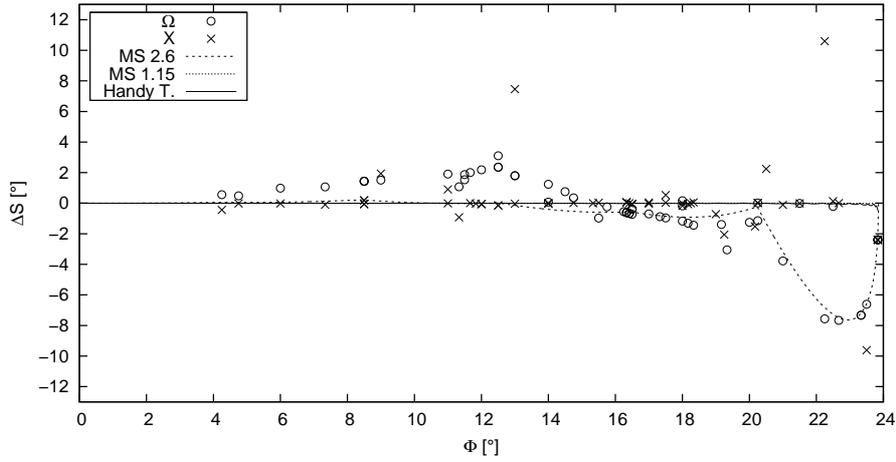}
\caption{Differences $\mathrm{\Delta}S=S-S(\mathrm{\Phi})$ between the distances $S(\mathrm{\Phi})$ from the summer solstice, trigonometrically computed from the latitude $\mathrm{\Phi}$ (Eq.~(\ref{eqn:S2B})), and the $S$-data of Book 8 ($\mathrm{\Omega}$-recension and X-manuscript) as well as $S$ resulting from an linear interpolation of the $S$-data of MS~1.15, MS~2.6, and Handy Tables ($\mathrm{\Delta}S \approx 0$ for MS~1.15 and Handy Tables)}
\label{fig:buch8_dL}
\end{center}
\end{figure}

\citet[p. 44]{stu06} mention that the $S$-data of X is considerably more precise that the $S$-data of $\mathrm{\Omega}$. Using a conversion based on Eq.~(\ref{eqn:S2B}) this can be confirmed.
In the case of $\mathrm{\Omega}$, the MAD for $\mathrm{\Phi}\mapsto S(\varepsilon_\mathrm{m})$ is $72\arcmin$. There are many differences $\mathrm{\Delta}S_i>1\arcdeg$ which cannot be explained by the resolution of the $S_i$. A wrong value for $\varepsilon$ underlying the reconversion can be excluded; the other (eligible) values do not lead to considerably  smaller coordinate differences.
Thus, either corrupt data must be assumed for $\mathrm{\Omega}$ or that the conversion done here is wrong.
Regarding the known sources of $S$-data, the tabulation in MS~1.15 has a step size of $1\arcdeg$ for $l$ and Ptolemy's Handy Tables include  a similar tabulation with a step size of $3\arcdeg$ \citep[p. 979]{neu75}. Thus, the usage of these tabulations (without interpolation) would result in $S$-values with fractions equal to zero. That does not hold for the data in Book 8.
In Fig.~\ref{fig:buch8_dL} the differences $\mathrm{\Delta} S_i = S_i-S(\mathrm{\Phi_i})$ are plotted against the latitude\footnote{\textit{Arbis} (8.25.14) with very large $|\mathrm{\Delta} S|$ is excluded in the case of $\mathrm{\Omega}$.}. The plot indicates a dependence between $\mathrm{\Delta}S$ and $\mathrm{\Phi}$ of $\mathrm{\Omega}$;  regionally  $\mathrm{\Delta}S$ increases or decreases with increasing $\mathrm{\Phi}$.
This suggests that the $S$-data in $\mathrm{\Omega}$ are not determined by means of the conversion (\ref{eqn:S2B}).

Table~\ref{tab:av_diff} also gives the MAD and the MD for the differences $\mathrm{\Delta\Phi}_i$ and $\mathrm{\Delta}S_i$ computed from the $\mathrm{\Phi}_i$ and converted $S(\mathrm{\Phi}_i)$ of the Canon. In the case of $\mathrm{\Omega}$ the results do not change significantly in comparison to the location catalogue of the Geography. In the case of X the Canon enables a comparison for all $S$-data (in contrast to the incomplete location catalogue of X).
Using the Canon, the small $\mathrm{MAD}=3.5\arcmin$ for $\mathrm{\Phi}\mapsto S(\varepsilon_\mathrm{m})$ confirms the high accuracy of the $S$-data which has already been observed for the small sample size of the location catalogue.
The large MD of $50\arcmin$ shows, as in case of the location catalogue, that there are also some latitudes which do not agree with the respective $S$.

Due to the incomplete location catalogue of X the missing latitudes were experimentally substituted by the values of $\mathrm{\Omega}$ as well as of the Canon to compute the $S(\mathrm{\Phi}_i)$. In the case of $\mathrm{\Omega}$ there are eight localities more than in the case of the Canon with large differences $|\mathrm{\Delta} S_i|>1\arcdeg$. \footnote{Localities with $|\mathrm{\Delta}S|>1\arcdeg$ occurring in $\mathrm{\Omega}$ but not in the Canon are: \textit{Kane} (8.22.9), \textit{Gerra} (8.22.10), \textit{Sapphara} (8.22.16), \textit{Chaberis} (8.26.5), \textit{Palura} (8.26.6), \textit{Patala} (8.26.10), \textit{Barbarei} (8.26.11), \textit{Tosalei} (8.27.5), \textit{Triglyphon} (8.27.7), and \textit{Aspithra} (8.27.11).}
Figure \ref{fig:buch8_dL} shows the $\mathrm{\Delta}S_i$ based on the latitudes of the Canon. In the case of X there is no comparable dependence between $\mathrm{\Delta}S$ and $\mathrm{\Phi}$ as detected for $\mathrm{\Omega}$. Most of the differences are near zero. Consequently, the coordinate differences of X reveal a close connection  between its $S$-data and the latitudes of the Canon.

Furthermore,  Fig.~\ref{fig:buch8_dL} shows the curves of the differences $\mathrm{\Delta}\tilde{S}(\mathrm{\Phi}) = \tilde{S}(\mathrm{\Phi})-S(\mathrm{\Phi})$ between the trigonometrically computed $S(\mathrm{\Phi})$ (Eq.~(\ref{eqn:S2B})) and the value $\tilde{S}(\mathrm{\Phi})$ determined by linear interpolations: 1) of the tabulation in MS~1.15, 2) of the tabulation in the Handy Tables\footnote{The tabulation of the Handy Tables is represented here by the data of the table in MS~1.15. To obtain a step size of $3\arcdeg$ for $l$ each third row was used.}, and 3) of the data of the parallels in MS~2.6.
The interpolations of the data in MS~1.15 and the Handy Tables lead to accurate results ($|\mathrm{\Delta}\tilde{S}|<2.3\arcmin$ and $|\mathrm{\Delta}\tilde{S}|<0.5\arcdeg$ respectively). Their error curves run close to zero so that they coincide with the majority of the differences $\mathrm{\Delta}S_i$ occurring in X. Consequently, a conversion by means of one of both tabulations can be assumed for the predominant amount of accurate $S_i$ in X. Because of the small number of data pairs $(\mathrm{\Phi}_j, S_j)$ in MS~2.6, the interpolation of these data  is significantly erroneous, especially for $\mathrm{\Phi}>20\arcdeg$. However, the curve $\mathrm{\Delta}\tilde{S}$ fits the differences $\mathrm{\Delta}S_i$ of some localities in  $\mathrm{\Omega}$ for $\mathrm{\Phi}>16\arcdeg$ well. This reveals the possibility of another type of interpolation underlying the data of $\mathrm{\Omega}$, see Sect.~\ref{sec:dss_appr}.

For the $S$-data of X and the auxiliary  latitudes of the Canon, Procedure \texttt{ROUND} (see Sect.~\ref{sec:alex}) was applied to explain as many $S_i$ as possible by rounding. Because of the small MAD, the standard resolution was set to $a_\mathrm{s}=\nicefrac{1}{8}\arcdeg=7.5\arcmin$. The less precise resolutions are $a=$ $\nicefrac{1}{6}\arcdeg$, $\nicefrac{1}{4}\arcdeg$, $\nicefrac{1}{3}\arcdeg$, $\nicefrac{1}{2}\arcdeg$, and $1\arcdeg$. As a result 42 of 59 $S_i$ can be explained by a rounding: $a=a_\mathrm{s}$: 15, $a=\nicefrac{1}{6}\arcdeg$: 20, $a=\nicefrac{1}{4}\arcdeg$: 3, $a=\nicefrac{1}{2}\arcdeg$: 3, and $a=1\arcdeg$: 1. Among the remaining 17 $S_i$ there are 5 with the obviously approximative value $S=0\arcdeg$\footnote{These localities are \textit{Syene} (8.15.15), \textit{Autolalai} (8.16.3), \textit{Gerra} (8.22.10), \textit{Harmuza} (8.22.21), and \textit{Tosalei} (8.27.5).}.
Furthermore, 2 $S_i$ can be explained by a linear interpolation of the data in MS~2.6, the concerning localities are \textit{Argyre} (8.27.10) and \textit{Kattigara} (8.27.13). Their $\mathrm{\Phi}$ is \ang{8;30}, the interpolated $S$ is \ang{68;49}, that is $\approx68\,\nicefrac{3}{4}\arcdeg$, the $S$ in Book 8.
The remaining 10 problematic $S_i$ probably have corruptions; for example, \textit{Dere} (8.16.12) has $\mathrm{\Phi}=\ang{11;00}$ and $S=62\,\nicefrac{3}{4}\arcdeg$, but the converted value is $S(\mathrm{\Phi})=\ang{61;51}\approx 61\,\nicefrac{3}{4}\arcdeg$ so that an error of $1\arcdeg$ in $S$ is possible.

\subsection{Linearly converted data in $\mathrm{\Omega}$} \label{sec:dss_appr}

\begin{figure}
\begin{center}
\includegraphics[width=120mm]{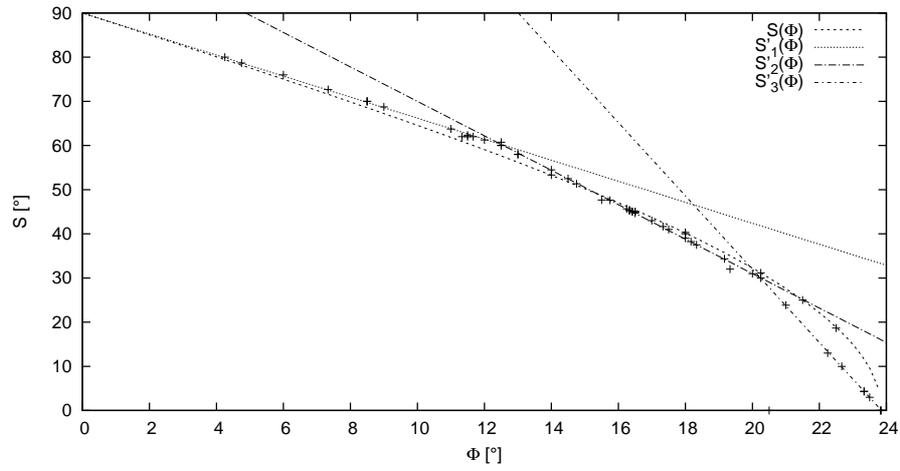}
\caption{$\mathrm{\Omega}$-recension: Plot of the distances $S_i$ from the summer solstice of Book 8 against the latitudes $\mathrm{\Phi}_i$ of the location catalogue, trigonometric conversion function $S(\mathrm{\Phi})$ (derived from Eq.~(\ref{eqn:S2B})), linear conversion functions $S'_1$, $S'_2$, and $S'_3$ (Eq.~(\ref{eqn:S2B_lin}))}
\label{fig:buch8_L}
\end{center}
\end{figure}

A few localities in Book 8 of $\mathrm{\Omega}$ have accurate $S$-values with $|\mathrm{\Delta}S|\leq 15\arcmin$, they are \textit{Geira} (8.16.6), \textit{Garama} (8.16.7), \textit{Napata} (8.16.8), \textit{Muza} (8.22.6), \textit{Karminna} (8.22.23), and \textit{Barbarei} (8.26.11).
The localities \textit{Syene} (8.15.15), \textit{Berenike} (8.15.19), and \textit{Autolalai} (8.16.3) have  $\mathrm{\Phi}=\ang{23;50}$ and the approximative $S=0\arcdeg$.
\textit{Arbis} (8.25.14) has the  erroneous $S=0\arcdeg$.
The $S$-values of most of the other localities can be explained by a conversion of $\mathrm{\Phi}$ by means of a piecewise linear  function.
The data reveal three regions for which a linear conversion was done separately. These regions cannot be made out well in the plot of the differences $\mathrm{\Delta}S_i$ in Fig.~ \ref{fig:buch8_dL} but by means of a plot of $S$ against $\mathrm{\Phi}$, see Fig.~ \ref{fig:buch8_L}.
Obviously, the limits between the three regions are $\mathrm{\Phi}_{12}\approx\ang{12;30}$ and $\mathrm{\Phi}_{23}\approx\ang{20;15}$; they correspond to a division of the quadrant of the ecliptic into three equal parts with the range of $\approx 30\arcdeg$.
Hence, the piecewise linear function of conversion is
\begin{equation} \label{eqn:S2B_lin}
S'(\mathrm{\Phi}) =  \begin{cases}
S'_1(\mathrm{\Phi}) = a_1 \mathrm{\Phi} + b_1 & \text{for }0\arcdeg\leq \mathrm{\Phi} < \mathrm{\Phi}_{12} \\
S'_2(\mathrm{\Phi}) = a_2 \mathrm{\Phi} + b_2 & \text{for }\mathrm{\Phi}_{12}\leq \mathrm{\Phi} < \mathrm{\Phi}_{23}\\
S'_3(\mathrm{\Phi}) = a_3 \mathrm{\Phi} + b_3 & \text{for }\mathrm{\Phi}_{23}\leq \mathrm{\Phi} < \varepsilon \quad \quad .
\end{cases}
\end{equation}

To define a linear functions two points are necessary. Two points $(\mathrm{\Phi},S)$ are known: $(0\arcdeg,90\arcdeg)$ and $(\varepsilon, 0\arcdeg)$. Moreover, very probably \textit{Badeo} (8.22.4) with $\mathrm{\Phi}_\mathrm{B}=\ang{20;15}$ and $S_\mathrm{B}=30\arcdeg$ was used as end point of $S'_1(\mathrm{\Phi})$ and starting point of $S'_2(\mathrm{\Phi})$ as well as \textit{Pityndra} (8.26.19) with $\mathrm{\Phi}_\mathrm{P}=\ang{12;30}$ and $S_\mathrm{P}=60\arcdeg$ was used as end point of $S'_2(\mathrm{\Phi})$ and starting point of $S'_3(\mathrm{\Phi})$. $S_\mathrm{B}$ and $S_\mathrm{P}$ are inaccurate, but conspicuously round values, possibly intensionally approximative to simplify the computations\footnote{An accurate $S$ for $\mathrm{\Phi}_\mathrm{B}$ exists in $\mathrm{\Omega}$.  \textit{Napata} (8.16.8) has $\mathrm{\Phi}=\ang{20;15}$ and $S=\ang{31;10}$, the value computed from Eq.~(\ref{eqn:S2B}) is $S(\mathrm{\Phi})=\ang{31;09}$.}.
It is obtained with $(0\arcdeg,90\arcdeg)$ and \textit{Pityndra}:
\begin{align}
a_1 &= (S_\mathrm{P}-90\arcdeg)/(\mathrm{\Phi}_\mathrm{P}-0\arcdeg) \approx - 2.4 \\
b_1 &= 90\arcdeg - a_1 \cdot 0\arcdeg = 90\arcdeg  \; .
\end{align}
It is obtained with \textit{Badeo} and  \textit{Pityndra}:
\begin{align}
a_2 &= (S_\mathrm{B}-S_\mathrm{P})/(\mathrm{\Phi}_\mathrm{B}-\mathrm{\Phi}_\mathrm{P}) \approx -3.9 \\
b_2 &= S_\mathrm{B} - a_2 \cdot \mathrm{\Phi}_\mathrm{B} = \ang{108;58} \approx 109\arcdeg \; .
\end{align}
It is obtained with \textit{Badeo} and $(\varepsilon, 0\arcdeg)$:
\begin{align}
a_3 &= (0\arcdeg-S_\mathrm{B})/(\varepsilon-\mathrm{\Phi}_\mathrm{B}) \approx -8.37 \sim -8.3 \\
b_3 &= 0\arcdeg - a_3 \cdot \varepsilon = \ang{197;49} \approx 198\arcdeg \; ,
\end{align}
where $\varepsilon=\ang{23;50}$ is used, the latitude of localities with $S=0\arcdeg$ in Book 8 (see above).

$S'(\mathrm{\Phi})$ is plotted in Fig.~\ref{fig:buch8_L} together with the given data points $(\mathrm{\Phi}_i, S_i)$ and the trigonometric conversion function $S(\mathrm{\Phi})$ based on Eq.~(\ref{eqn:S2B}). $S'$ fits the inaccurate data points very well. Especially $S'_3$ is a bad approximation for the trigonometric function $S(\mathrm{\Phi})$ based on Eq.~(\ref{eqn:S2B}).
In Table~\ref{tab:ass} $S_i$, the computed $S'(\mathrm{\Phi}_i)$ as well as the remaining difference $\mathrm{\Delta}S'_i=S_i-S'(\mathrm{\Phi}_i)$ are given for the 47 localities which have not been discussed so far.
The differences $\mathrm{\Delta}S'_i$ are explicable in almost every case by a reasonable rounding to the nearest $1\arcdeg$, $\nicefrac{1}{2}\arcdeg$, $\nicefrac{1}{3}\arcdeg$, or $\nicefrac{1}{4}\arcdeg$. There remain 10 localities whose $S'(\mathrm{\Phi})$ cannot be rounded to the given $S$ in this way. The reason could be peculiar roundings or scribal errors, also in an originally trigonometric converted $S$.
For example, a linear conversion and a confusion of $\nicefrac{1}{3}=\gamma$ and $\nicefrac{2}{3}=\gamma o$ can be assumed for \textit{Barygaza} ($S'(\mathrm{\Phi})=\ang{41;24}\approx 41\,\nicefrac{1}{3}\arcdeg\neq 41\,\nicefrac{2}{3}\arcdeg = S$) as well as for \textit{Maagrammon}.
A trigonometric conversion and a scribal error can be assumed for \textit{Iazeitha}, which has  $S=\ang{47;40}$ in $\mathrm{\Omega}$ and the accurate value $\ang{48;40}$ in X.

\begin{appendix}

\section*{Appendix: Changes to and errors in the data source}  %\label{sec:db_bern}

The following changes of coordinates values underlie the present investigations.
Changes of known inconsistent values are:
$\mathrm{\Phi}$ of \textit{Autolalai} (4.6.24) in X was changed from \ang{27;20} to  $\ang{23;50}$ (as $\mathrm{\Omega}$ and Canon). \textit{Argyre} (8.27.10) and \textit{Kattigara} (8.27.13) were positioned on the northern hemisphere instead of the southern one. (The positive latitudes equal the more plausible values in the Canon. On the negative latitudes of Chinese places see also \citet[pp. 47]{raw08}.) Further changes are: \textit{Palmyra} (8.20.10): $M = 14^{\mathrm{h}}40^{\mathrm{m}}$ \citep[also][]{nob66} to $M = 14^{\mathrm{h}}20^{\mathrm{m}}$ \citep{raw09} and \textit{Apameia} (8.20.9):  $A=\nicefrac{1}{8}$\,h to \nicefrac{5}{8}\,h \citep{nob66, raw09}.

As a support for further investigations of the coordinates of Book 8, the errors shall be given here, which were found in the data base of \cite{stu06}. Data which are correct in the printout but erroneous in the data base are given in Table~\ref{tab:errors_stu06}.

\end{appendix}

\bibliographystyle{natdin}
\setlength{\bibsep}{1mm}

%%%%%%%%%%%%%%%%%%%%%%%%%%%%%%%%%%%%%%%%%%%%%%%%%%%%%%%%%%%%%%%%%%%
%                            TABLES
%%%%%%%%%%%%%%%%%%%%%%%%%%%%%%%%%%%%%%%%%%%%%%%%%%%%%%%%%%%%%%%%%%%

\clearpage

% 1
\begin{table}
\centering
\caption{Median absolute deviation MAD (Eq.~(\ref{eqn:mad})) and mean deviation MD (Eq.~(\ref{eqn:md})) after conversions of coordinates of the $\mathrm{\Omega}$-recension, the X-manuscript, and the Canon;
$M\mapsto \mathrm{\Phi}$: deviations $\mathrm{\Phi}_i-\mathrm{\Phi}(M_i)$ between the latitudes $\mathrm{\Phi}(M_i)$, computed from the lengths of the longest day $M_i$ of Book 8, and the $\mathrm{\Phi}_i$ of Books 2--7/the Canon;
$\mathrm{\Phi}\mapsto M$: deviations $M_i-M(\mathrm{\Phi}_i)$ between the $M(\mathrm{\Phi}_i)$, computed from the $\mathrm{\Phi}_i$ of Books 2--7/the Canon, and the $M_i$ of Book 8;
$S\mapsto\mathrm{\Phi}$ and $\mathrm{\Phi}\mapsto S$: as $M\mapsto \mathrm{\Phi}$ and $\mathrm{\Phi}\mapsto M$ but for the distances $S_i$ of the sun from the summer solstice of Book 8;
$n$ is the number of coordinates;
the values of the obliquity of the ecliptic underlying the conversions are: $\varepsilon_\mathrm{m} = \ang{23;51;20}, \varepsilon_\mathrm{h} = \ang{23;55}, \varepsilon_\mathrm{r} = 24\arcdeg, \varepsilon_\mathrm{c} = \ang{23;41}$}
\label{tab:av_diff}
\begin{tabular} {llrrrrrrrr}
\noalign{\smallskip}\noalign{\smallskip}\toprule
  \multicolumn{2}{l}{Conversion} & \multicolumn{2}{c}{$\mathrm{\Omega}$ B.2--7/B.8} & \multicolumn{2}{c}{Canon/$\mathrm{\Omega}$ B.8} & \multicolumn{2}{c}{X B.2--5/B.8} & \multicolumn{2}{c}{Canon/X B.8}\\
\cmidrule(lr){3-4} \cmidrule(lr){5-6} \cmidrule(lr){7-8} \cmidrule(lr){9-10}
  & $\varepsilon$ & \multicolumn{1}{c}{MAD} & \multicolumn{1}{c}{MD} & \multicolumn{1}{c}{MAD} & \multicolumn{1}{c}{MD} & \multicolumn{1}{c}{MAD} & \multicolumn{1}{c}{MD} & \multicolumn{1}{c}{MAD} & \multicolumn{1}{c}{MD}\\
\midrule
 \multicolumn{2}{l}{$n$} & 344 & 344 & 346 & 346 & 220 & 220 & 354 & 354 \\
\midrule
$M\mapsto \mathrm{\Phi}$ & $\varepsilon_\mathrm{m}$ & 9.7\arcmin & 13.0\arcmin & 10.7\arcmin & 17.5\arcmin & 9.6\arcmin & 13.6\arcmin & 10.7\arcmin & 19.8\arcmin\\
 & $\varepsilon_\mathrm{h}$ & 7.9\arcmin & 13.6\arcmin & 11.6\arcmin & 18.1\arcmin & 8.5\arcmin & 14.4\arcmin & 10.7\arcmin & 20.3\arcmin\\
 & $\varepsilon_\mathrm{r}$ & 11.4\arcmin & 15.9\arcmin & 12.8\arcmin & 20.1\arcmin & 12.0\arcmin & 17.3\arcmin & 11.8\arcmin & 22.2\arcmin\\
 & $\varepsilon_\mathrm{c}$ & 14.3\arcmin & 18.4\arcmin & 15.2\arcmin & 22.3\arcmin & 15.4\arcmin & 18.6\arcmin & 15.8\arcmin & 25.1\arcmin\\
\midrule
$\mathrm{\Phi}\mapsto M$ & $\varepsilon_\mathrm{m}$ & 0.9$^{\text{m}}$ & 1.3$^{\text{m}}$ & 1.0$^{\text{m}}$ & 1.7$^{\text{m}}$ & 1.0$^{\text{m}}$ & 1.5$^{\text{m}}$ & 1.1$^{\text{m}}$ & 1.9$^{\text{m}}$\\
   & $\varepsilon_\mathrm{h}$ & 0.9$^{\text{m}}$ & 1.4$^{\text{m}}$ & 1.2$^{\text{m}}$ & 1.8$^{\text{m}}$ & 1.1$^{\text{m}}$ & 1.6$^{\text{m}}$ & 1.1$^{\text{m}}$ & 2.0$^{\text{m}}$\\
 & $\varepsilon_\mathrm{r}$ & 1.2$^{\text{m}}$ & 1.6$^{\text{m}}$ & 1.5$^{\text{m}}$ & 2.1$^{\text{m}}$ & 1.5$^{\text{m}}$ & 2.0$^{\text{m}}$ & 1.4$^{\text{m}}$ & 2.2$^{\text{m}}$\\
 & $\varepsilon_\mathrm{c}$ & 1.5$^{\text{m}}$ & 1.9$^{\text{m}}$ & 1.6$^{\text{m}}$ & 2.3$^{\text{m}}$ & 1.8$^{\text{m}}$ & 2.1$^{\text{m}}$ & 1.7$^{\text{m}}$ & 2.5$^{\text{m}}$\\
\midrule
 \multicolumn{2}{l}{$n$} & 58 & 58 & 59 & 59 & 13 & 13 & 58 & 58 \\
\midrule
$S\mapsto\mathrm{\Phi}$ & $\varepsilon_\mathrm{m}$ & 18.7\arcmin & 22.0\arcmin & 19.3\arcmin & 27.3\arcmin & 0.5\arcmin & 3.5\arcmin & 1.1\arcmin & 8.6\arcmin\\
& $\varepsilon_\mathrm{h}$ & 21.2\arcmin & 23.1\arcmin & 21.7\arcmin & 28.0\arcmin & 3.2\arcmin & 5.4\arcmin & 3.1\arcmin & 10.2\arcmin\\
& $\varepsilon_\mathrm{r}$ & 22.4\arcmin & 24.8\arcmin & 22.4\arcmin & 29.4\arcmin & 6.9\arcmin & 8.1\arcmin & 6.6\arcmin & 12.6\arcmin\\
& $\varepsilon_\mathrm{c}$ & 13.1\arcmin & 21.0\arcmin & 15.0\arcmin & 27.3\arcmin & 7.5\arcmin & 9.5\arcmin & 7.5\arcmin & 12.9\arcmin\\
\midrule
$\mathrm{\Phi}\mapsto S$ & $\varepsilon_\mathrm{m}$ & 72.4\arcmin & 104.6\arcmin & 86.2\arcmin & 117.5\arcmin & 1.5\arcmin & 31.2\arcmin & 3.5\arcmin & 50.0\arcmin\\
 & $\varepsilon_\mathrm{h}$ & 81.5\arcmin & 117.4\arcmin & 82.7\arcmin & 124.1\arcmin & 11.6\arcmin & 57.1\arcmin & 11.3\arcmin & 64.3\arcmin\\
 & $\varepsilon_\mathrm{r}$ & 78.6\arcmin & 131.7\arcmin & 79.6\arcmin & 138.6\arcmin & 23.5\arcmin & 82.2\arcmin & 21.8\arcmin & 80.4\arcmin\\
\midrule
 \multicolumn{2}{l}{$n$} & 55 & 55 & 54 & 54 & 11 & 11 & 54 & 54 \\
\midrule
 & $\varepsilon_\mathrm{c}$ & 57.4\arcmin & 90.1\arcmin & 72.7\arcmin & 110.3\arcmin & 23.0\arcmin & 32.2\arcmin & 22.2\arcmin & 57.1\arcmin\\
\bottomrule
\end{tabular}
\end{table}

% 2
\begin{table}													
\centering													
\caption{Examples of the conversion of the latitude $\mathrm{\Phi}$ of the location catalogue into the length of the longest day $M(\mathrm{\Phi})$ by Eq.~(\ref{eqn:M2B}) with $\varepsilon=\varepsilon_\mathrm{m}$; $M$ is the coordinate value of Book 8} \label{tab:raw}													
\begin{tabular} { l l l l l l}
\noalign{\smallskip}\noalign{\smallskip}\toprule
\multicolumn{1}{l}{Locality}
& \multicolumn{1}{c}{$\mathrm{\Phi}$}
& \multicolumn{2}{c}{$M$}
& \multicolumn{1}{c}{$M(\mathrm{\Phi}$)} \\
\cmidrule(lr){3-4}
& \multicolumn{1}{c}{[\arcdeg]}	
& \multicolumn{1}{c}{[h]}
& \multicolumn{1}{c}{[h,min]}
& \multicolumn{1}{c}{[h,min]}\\
\midrule
\textit{Arbela} (8.21.3)  & 37\,\nicefrac{1}{4}	& 14\,\nicefrac{5}{8} & 14,17.5 & 14,17.2 \\
\textit{Athenai} (8.12.18)& 37\,\nicefrac{1}{4}	& 14\,\nicefrac{5}{8} & 14,17.5 & 14,17.2 \\
\textit{Karchedon}  (8.14.5)& 32\,\nicefrac{2}{3} & 14\,\nicefrac{1}{5} & 14,12.0 & 14,11.8 \\
\textit{Lysimachia}  (8.11.7)& 41\,\nicefrac{1}{2} 	 &15\,\nicefrac{1}{12} & 15,05.0 &  15,04.3 \\
\textit{Nikaia}  (8.17.7)& 41\,\nicefrac{11}{12}	&	 15\,\nicefrac{1}{8} & 15,07.5 & 15,07.1 \\
\textit{Rhodos}  (8.17.21)& 36      	&14\,\nicefrac{1}{2} & 14,30.0 & 14,29.9 \\
\textit{Tarentum}  (8.8.4)& 40 & 14\,\nicefrac{11}{12} & 14,55.0 &  14,54.2 \\
\bottomrule
\end{tabular}													
\end{table}

% 3
\begin{table}													
\centering													
\caption{Estimated percentages of different resolutions $a$ of the lengths of the longest day $M$ of Book 8 ($\mathrm{\Omega}$-recension and X-manuscript) in the case that they originate from a trigonometric conversion (Eq.~(\ref{eqn:M2B})) or from a linear interpolation of the $M$-data in MS 2.6 and GH 1.23; last row: $M$ does not fit with $\mathrm{\Phi}$} \label{tab:aufloes_T}													 \begin{tabular} {l r r  r r  r r  r}
\noalign{\smallskip}\noalign{\smallskip}\toprule
& \multicolumn{3}{c}{$\mathrm{\Omega}$}
& \multicolumn{3}{c}{X} \\
\cmidrule(lr){2-4} \cmidrule(lr){5-7}
& \multicolumn{1}{c}{trig.}
& \multicolumn{1}{c}{interp.}
& \multicolumn{1}{c}{interp.}
& \multicolumn{1}{c}{trig.}
& \multicolumn{1}{c}{interp.}
& \multicolumn{1}{c}{interp.} \\
&
& \multicolumn{1}{c}{MS}
& \multicolumn{1}{c}{GH}
&
& \multicolumn{1}{c}{MS}
& \multicolumn{1}{c}{GH} \\
\midrule
$a<\nicefrac{1}{12}$\,h & 12 & 12 & 11 & 3 & 3 & 2 \\
$a=\nicefrac{1}{12}$\,h & 72 & 72 & 68 & 71 & 69 & 65 \\
$a>\nicefrac{1}{12}$\,h & 12 & 12 & 18 & 21 & 22 & 24 \\
\multicolumn{1}{c}{$M \ncong M(\mathrm{\Phi})$} & 4 & 3 & 4 & 5 &  6 & 9  \\
\bottomrule
\end{tabular}													
\end{table}

\clearpage

% 4
\begin{longtable}
{l
 >{$}r<{$}
 >{$}r<{$}
 >{$}r<{$}
 >{$}r<{$}}
\caption{$\mathrm{\Omega}$-recension: Linear conversion of the latitude $\mathrm{\Phi}$ of the location catalogue into the distance $S'(\mathrm{\Phi})$ from the summer solstice by means of the piecewise functions $S'_1(\mathrm{\Phi})$, $S'_2(\mathrm{\Phi})$, and $S'_3(\mathrm{\Phi})$ (Eq.~(\ref{eqn:S2B_lin})) and remaining difference $\mathrm{\Delta}S'=S-S'(\mathrm{\Phi})$ with respect to $S$ of Book 8}
\label{tab:ass}\\
\noalign{\smallskip}\noalign{\smallskip}\toprule
 \multicolumn{1}{l}{Locality} &
 \multicolumn{1}{c}{$\mathrm{\Phi}$} &
 \multicolumn{1}{c}{$S$} &
 \multicolumn{1}{c}{$S'$} &
 \multicolumn{1}{c}{$\mathrm{\Delta}S'$} \\
  &
 \multicolumn{1}{c}{[\arcdeg,\arcmin]} &
 \multicolumn{1}{c}{[\arcdeg,\arcmin]} &
 \multicolumn{1}{c}{[\arcdeg,\arcmin]} &
 \multicolumn{1}{c}{[\arcmin]} \\
\midrule
\endfirsthead
\noalign{\smallskip}\noalign{\smallskip}\toprule
 \multicolumn{1}{l}{Locality} &
 \multicolumn{1}{c}{$\mathrm{\Phi}$} &
 \multicolumn{1}{c}{$S$} &
 \multicolumn{1}{c}{$S'$} &
 \multicolumn{1}{c}{$\mathrm{\Delta}S'$} \\
  &
 \multicolumn{1}{c}{[\arcdeg,\arcmin]} &
 \multicolumn{1}{c}{[\arcdeg,\arcmin]} &
 \multicolumn{1}{c}{[\arcdeg,\arcmin]} &
 \multicolumn{1}{c}{[\arcmin]} \\
\midrule
\endhead
 \multicolumn{5}{l}{$S'_1(\mathrm{\Phi})$} \\
\textit{Adulis} (16.11) &  11,20 &  62,00 &  62,48 & -48 \\
\textit{Dere} (16.12) &  11,00 &  63,45 &  63,36 & 9 \\
\textit{Mosylon} (16.13) &  9,00 &  68,45 &  68,24 & 21 \\
\textit{Aromata} (16.14) &  6,00 &  76,00 &  75,36 & 24 \\
\textit{Okelis} (22.7) &  12,00 &  61,15 &  61,12 & 3 \\
\textit{Arabia} (22.8) &  11,30 &  62,20 &  62,24 & -4 \\
\textit{Palura} (26.6) &  11,30 &  62,00 &  62,24 & -24 \\
\textit{Takola} (27.3) &  4,15 &  80,00 &  79,48 & 12 \\
\textit{Zabai} (27.4) &  4,45 &  78,40 &  78,36 & 4 \\
\textit{Argyre} (27.10) &  8,30 &  70,00 &  69,36 & 24 \\
\textit{Kattigara} (27.13) &  8,30 &  70,00 &  69,36 & 24 \\
\textit{Nagadiba} (28.3) &  8,30 &  70,00 &  69,36 & 24 \\
\textit{Talakori} (28.4) &  11,40 &  62,00 &  62,00 & 0 \\
\textit{Maagrammon} (28.5) &  7,20 &  72,40 &  72,24 & 16 \\
\midrule
 \multicolumn{5}{l}{$S'_2(\mathrm{\Phi})$} \\
 \textit{Iarzeitha} (16.4) &  15,30 &  47,40 &  48,33 & -53 \\
 \textit{Thamondokana} (16.5) &  17,00 &  43,00 &  42,42 & 18 \\
 \textit{Meroë} (16.9) &  16,25 &  45,00 &  44,58 & 2 \\
 \textit{Ptolemaïs Theron} (16.10) &  16,25 &  45,00 &  44,58 & 2 \\
 \textit{Town of Pudnos} (22.5) &  16,30 &  44,40 &  44,39 & 1 \\
 \textit{Kane} (22.9) &  12,30 &  60,45 &  60,15 & 30 \\
 \textit{Mara} (22.11) &  18,20 &  37,30 &  37,30 & 0 \\
 \textit{Omanon} (22.12) &  19,20 &  32,00 &  33,36 & -96 \\
 \textit{Menambis} (22.13) &  16,30 &  45,00 &  44,39 & 21 \\
 \textit{Sabbatha} (22.14) &  16,30 &  45,00 &  44,39 & 21 \\
 \textit{Sabe} (22.15) &  13,00 &  58,00 &  58,18 & -18 \\
 \textit{Sapphara} (22.16) &  14,30 &  52,30 &  52,27 & 3 \\
 \textit{Isle of Sarapias} (22.18) &  17,30 &  41,00 &  40,45 & 15 \\
 \textit{Simylla} (26.3) &  14,45 &  51,20 &  51,29 & -8 \\
 \textit{Muziris} (26.4) &  14,00 &  54,30 &  54,24 & 6 \\
 \textit{Chaberis} (26.5) &  15,45 &  47,36 &  47,35 & 1 \\
 \textit{Barygaza} (26.12) &  17,20 &  41,40 &  41,24 & 16 \\
 \textit{Ozene} (26.13) &  20,00 &  31,00 &  31,00 & 0 \\
 \textit{Baithana} (26.14) &  18,10 &  38,15 &  38,09 & 6 \\
 \textit{Hippokura} (26.15) &  19,10 &  34,20 &  34,15 & 5 \\
 \textit{Karura} (26.16) &  16,20 &  45,20 &  45,18 & 2 \\
 \textit{Modura} (26.17) &  16,20 &  45,20 &  45,18 & 2 \\
 \textit{Orthura} (26.18) &  16,20 &  45,20 &  45,18 & 2 \\
 \textit{Triglyphon} (27.7) &  18,00 &  39,00 &  38,48 & 12 \\
 \textit{Marëura} (27.8) &  12,30 &  60,00 &  60,15 & -15 \\
 \textit{Aspithra} (27.11) &  16,15 &  45,40 &  45,38 & 2 \\
 \textit{Sinai} (27.12) &  13,00 &  58,00 &  58,18 & -18 \\
\midrule
 \multicolumn{5}{l}{$S'_3(\mathrm{\Phi})$} \\
 \textit{Gerra} (22.10) &  23,20 &  4,20 &  4,20 & 0 \\
 \textit{Harmuza} (22.21) &  23,30 &  3,00 &  2,57 & 3 \\
 \textit{Samydake} (22.22) &  22,40 &  10,00 &  9,52 & 8 \\
 \textit{Patala} (26.10) &  21,00 &  23,50 &  23,42 & 8 \\
 \textit{Tosalei} (27.5) &  23,20 &  4,20 &  4,20 & 0 \\
 \textit{Tugma} (27.6) &  22,15 &  13,00 &  13,19 & -19 \\
\noalign{\smallskip}\hline % \bottomrule
\end{longtable}

% 5
\vspace{2cm}
\begin{table} [h]
\centering
\caption{Errors in the data base of \cite{stu06}}
\label{tab:errors_stu06}
\begin{tabular}{lllll}
\noalign{\smallskip}\noalign{\smallskip}\toprule
ID & Name & $\mathrm{\Omega}$/X& Error & Correction\\
\midrule
8.3.7.1 & \textit{Eboracum} & X & link to 2.3.16 & 2.3.17 \\
8.5.7.3 & \textit{Narbo} & X &  $M = 16^{\mathrm{h}}15^{\mathrm{m}}$ &  $15^{\mathrm{h}}15^{\mathrm{m}}$ \\
8.5.7.4 & \textit{Vienna} & $\mathrm{\Omega}$ &  $A =2^{\mathrm{h}}30^{\mathrm{m}}$ & $2^{\mathrm{h}}15^{\mathrm{m}}$\\
8.5.7.4 & \textit{Vienna} & X & $A = 2^{\mathrm{h}}15^{\mathrm{m}}$ & $2^{\mathrm{h}}30^{\mathrm{m}}$ \\
8.5.7.4 & \textit{Vienna} & X & $M = 17^{\mathrm{h}}15^{\mathrm{m}}$ & $15^{\mathrm{h}}30^{\mathrm{m}}$ \\
8.8.6.1 & \textit{Aquileia} & $\mathrm{\Omega}$, X & $M=15^{\mathrm{h}}05^{\mathrm{m}}$ & $15^{\mathrm{h}}30^{\mathrm{m}}$  \\
8.13.8.1 & \textit{Iol Caesarea} & X & $M = 14^{\mathrm{h}}20^{\mathrm{m}}$ & $14^{\mathrm{h}}15^{\mathrm{m}}$ \\
8.17.32.1 & \textit{Perge/Aksu} & X &  $A = 0^{\mathrm{h}}30^{\mathrm{m}}$ & $0^{\mathrm{h}}05^{\mathrm{m}}$\\
8.17.46.1 & \textit{Adana} & $\mathrm{\Omega}$ & link to 5.8.8 & 5.8.7 \\
8.23.12.1 & \textit{Maruka} & $\mathrm{\Omega}$ & $A = \nicefrac{35}{6}\,\mathrm{h}$ & $3\,\nicefrac{5}{6}\,\mathrm{h}$ \\
8.25.7.1 & \textit{Ortospana} & X &  $M=15^{\mathrm{h}}20^{\mathrm{m}}$ & $14^{\mathrm{h}}20^{\mathrm{m}}$ \\
8.27.7.1 & \textit{Trilingon} & $\mathrm{\Omega}$ & $A=7^{\mathrm{h}}16^{\mathrm{m}}$ & $6^{\mathrm{h}}16^{\mathrm{m}}$ \\
\bottomrule
\end{tabular}
\end{table}

%%%%%%%%%%%%%%%%%%%%%%%%%%%%%%%%%%%%%%%%%%%%%%%%%%%%%%%%%%%%%%%%%%%
%                            FIGURES
%%%%%%%%%%%%%%%%%%%%%%%%%%%%%%%%%%%%%%%%%%%%%%%%%%%%%%%%%%%%%%%%%%%
\clearpage

% 1

% 2

% 3

% 4

% 5

% 6

% 7

% 8

% 9

% 10

% 11

\end{document}